\documentclass[prb,aps,a4paper,twocolumn]{revtex4}

\usepackage{float}
\usepackage{pst-node}
\usepackage{amsmath}
\usepackage{graphicx}
\usepackage{latexsym}
\usepackage{amsfonts}
\usepackage{amssymb}
\usepackage{bm}
\usepackage{graphicx}
\usepackage{subfig}
\usepackage{xfrac}
\usepackage{multirow}

\usepackage{rotating}
\usepackage{cancel}
\usepackage[nouppercase]{scrpage2}
\usepackage{verbatim}
\usepackage{color}

\usepackage[colorlinks=true,bookmarks=false,pdfstartview={XYZ null null 1}]{hyperref} %PdfStart X=posX Y=posY Z=zoom
\usepackage[all]{hypcap} %Hyperlinking references
\usepackage{cleveref} %Reference stuff with \cref or \Cref (capitalised)

\begin{document}
\title{\Large
{\bf Excitons in van der Waals heterostructures: \\
The important role of dielectric screening}
\vspace{0.5cm}}
\author{\bf S. Latini, T. Olsen, K.S. Thygesen
\vspace{0.2cm}}

\affiliation{Center for Nanostructured Graphene (CNG) and Center for Atomic-scale Materials Design (CAMD), Department of Physics, Technical University of Denmark, 2800 Kgs. Lyngby, Denmark }

\begin{abstract}

The existence of strongly bound excitons is one of the hallmarks of the newly discovered atomically thin semi-conductors. While it is understood that the large binding energy is mainly due to the weak dielectric screening in two dimensions (2D), a systematic investigation of the role of screening on 2D excitons is still lacking. Here we provide a critical assessment of a widely used 2D hydrogenic exciton model which assumes a dielectric function of the form $\epsilon(q)=1+2\pi\alpha q$, and we develop a quasi-2D model with a much broader applicability. Within the quasi-2D picture, electrons and holes are described as in-plane point charges with a finite extension in the perpendicular direction and their interaction is screened by a dielectric function with a non-linear $q$-dependence which is computed \emph{ab-initio}. The screened interaction is used in a generalized Mott-Wannier model to calculate exciton binding energies in both isolated and supported 2D materials. For isolated 2D materials, the quasi-2D treatment yields results almost identical to those of the strict 2D model and both are in good agreement with \emph{ab-initio} many-body calculations. On the other hand, for more complex structures such as supported layers or layers embedded in a van der Waals heterostructure, the size of the exciton in reciprocal space extends well beyond the linear regime of the dielectric function and a quasi-2D description has to replace the 2D one. Our methodology has the merit of providing a seamless connection between the strict 2D limit of isolated monolayer materials and the more bulk-like screening characteristics of supported 2D materials or van der Waals heterostructures.
\end{abstract}
\maketitle

\vspace{1cm}

\section{Introduction}
Atomically thin semiconductors\cite{Wang2012} like graphene, hexagonal boron-nitride (hBN), and MoS$_2$ are presently being intensely studied due to their extraordinary opto-electronic properties. It is characteristic for these two-dimensional (2D) semiconductors 
that excitonic effects play a fundamental role, substantially modifying the optical spectrum by introducing states within the band gap that couple strongly to light and shift the onset of optical transitions to lower energies.\cite{Fai2010,Splendiani2010,Ramasubramaniam2012,Qiu2013,
Ugeda2014,Keliang2014}. Knowledge of the nature of the excitonic states is thus essential for device engineering\cite{Jariwala2014,Bernardi2013,Lopez2013,Ross2014,Pospischil2014}. The well known Mott-Wannier model\cite{GrossoPastori2000}, which schematizes the exciton as a bound electron-hole pair interacting via a statically screened Coulomb interaction, is widely used to estimate exciton binding energies and radii in bulk materials. The main approximations behind the Mott-Wannier model are essentially three: (i) The real band structure is replaced by two parabolic bands. (ii) The microscopic shape of the conduction and valence band wave functions is neglected. (iii) The dielectric function is assumed to be local in real space, i.e $q$-independent in reciprocal space. For 2D materials, the performance of the Mott-Wannier model and the validity of the underlying approximations have still not been systematically investigated. The present work focuses on (iii), which is the only approximation where the role of the reduced dimensionality represents a qualitative difference from the 3D case.

For bulk semiconductors the macroscopic dielectric constant is defined as the limiting value of $\epsilon (q)$ as $q\rightarrow0$. For a 2D semiconductor this definition cannot be straightforwardly adopted since $\epsilon(q=0)=1$. In fact, for 2D systems the dielectric function is strongly $q$-dependent and a more elaborate treatment of screening is required\cite{Cudazzo2010,Cudazzo2011,Falco2013}. This issue has been treated by several authors\cite{Cudazzo2010,Cudazzo2011,Pulci2012,Hybertsen2013}, who envisioned the 2D material as a strict 2D system, i.e. mathematically 2D, with a dielectric function of the form
\begin{equation}
\label{eq:lineps}
\epsilon_{2D}({\bf q}) = 1+2\pi\alpha q,
\end{equation}
where $\alpha$ is the 2D polarizability of the layer, which can be computed \emph{ab-initio}. The screened electron-hole interaction then follows
\begin{equation}
\label{eq:2Dpotq}
W_{2D}({\bf q}) = -\frac{2\pi}{q}\epsilon_{2D}^{-1}(q),
\end{equation}
where $2\pi/q$ is the 2D Fourier transform of $1/r$. This form of interaction has the merit of leading to an analytical potential in real space and it has been successfully used to describe exciton binding energies and radii of several 2D systems \cite{Cudazzo2011,Pulci2012,Hybertsen2013}. We note that the form $1/q$ for the interaction and Eq. (\ref{eq:lineps}) for the dielectric function are consistent approximations which both become exact in the limit of vanishing thickness of the material, i.e. the strict 2D limit. However, to the best of our knowledge the validity range and limitations of these approximations have not previously been systematically explored. 

In this paper we relax the assumptions behind the 2D model adopting a microscopic approach that accounts for both the finite thickness of the layer and the full wave-vector dependence of the dielectric function. In the case of isolated monolayers our quasi-2D (Q2D) description agrees well with the established strict 2D model providing a justification for the latter. However, in the case of 2D layers supported on semi-infinite substrates or for thicker, i.e. few-layer, 2D materials, we find it important to account for the finite thickness and include the full non-linear $q$-dependence of the dielectric function. In a recent paper we introduced a method for calculating the dielectric function of general layered materials (so-called van der Waals heterostructures\cite{Terrones2013,Britnell2013,Geim2013}) where the dielectric functions of the individual layers are computed \emph{ab-initio} and subsequently coupled together electrostatically\cite{Andersen2015}. In the present work we use this method to compute the screened electron-hole interaction and solve the resulting quasi-2D Mott-Wannier model for various types of heterostructures. We show that the exciton binding energy and radius can be effectively tuned by controlling the screening via the heterostructure environment. Surprisingly we find that the transition from a strongly bound exciton in monolayer MoS$_2$ (binding energy of 0.6 eV) to a weakly bound exciton in bulk MoS$_2$ (binding energy of 0.15 eV) can be seamlessly described by the quasi-2D Mott-Wannier model accounting only for the change in the screening. 

%The paper is organized as follows. We start by motivating the quasi-2D picture and deriving an expression for an effective unscreened electron-hole interaction. We then %screen the interaction by introducing a macroscopic wave-vector dependent dielectric function, which can be calculated ab-initio. To illustrate the validity of the %method we calculate the energy of the lowest bound exciton in hBN and MoS$_2$ using our screened interaction in a generalized Mott-Wannier model and benchmark the %results against fully converged values obtained from the ab-initio solution of Bethe Salpeter Equation (BSE)\cite{Strinati1984,Onida2002}. We further report that for %isolated layers the Q2D and 2D binding energies coincide, justifying a strict 2D treatment from a microscopic point of view. In the last part we show that the 2D model %breaks down when dealing with stacks of 2D semiconductors, i.e. van der Waals heterostructures or 2D materials on a substrate. In this case accounting for the full %wave-vector dependence of the screening turns out to be essential. \\

\section{The quasi-2D picture}
Even though atomically thin semiconductors are referred to as 2D materials they obviously do have a finite thickness. In this section, we show how the finite thickness can be accounted for within a 2D description. We shall refer to this description as the quasi-2D picture. To illustrate the concept, we consider the interaction between two arbitrary charge distributions. 
\begin{figure}[t]
\centering  
\includegraphics[width=0.43\textwidth]{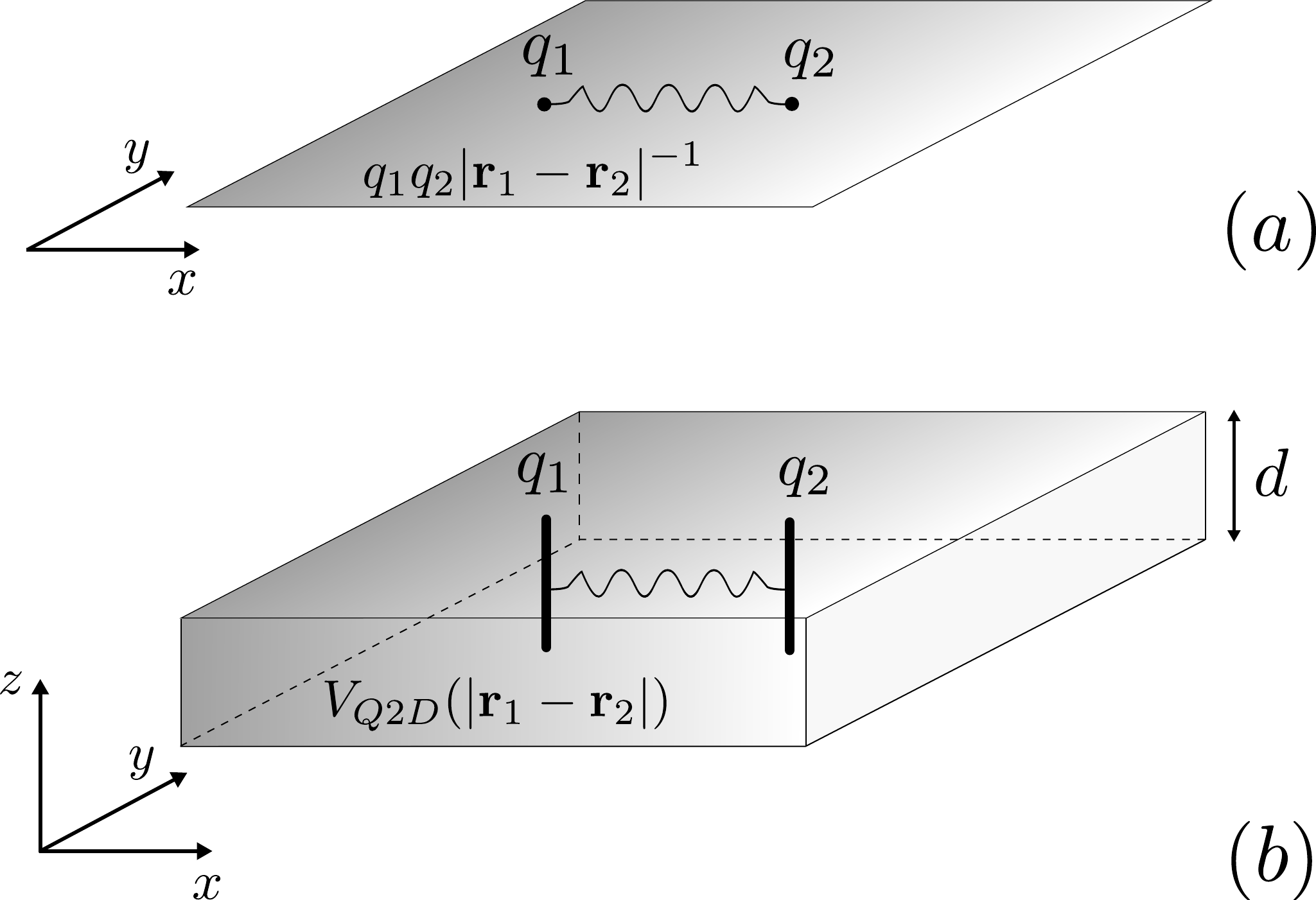}
\caption{Sketch of the (a) pure 2D and (b) quasi-2D Coulomb interaction. In the latter case the point charges can be thought as lines of charge extending along the thickness of the material. }
\label{fig:InteractionSketch}
\end{figure}
\begin{equation}
\label{eq:bareVgeneral}
V_{12} = \int d{\bf r}d{\bf r}'\frac{\rho_1({\bf r})\rho_2({\bf r}')}{|{\bf r}-{\bf r}'|}.
\end{equation}

In the case of two point charges confined to a 2D plane (see \cref{fig:InteractionSketch}(a)), each charge distribution is given by a delta function, i.e. $\rho_i({\bf r}_\parallel)=q_i\delta({\bf r}_\parallel-{\bf r}_{i,\parallel})$, leading to an interaction in reciprocal space:
\begin{equation}
V_{2D}({\bf q}_\parallel) =q_1q_2\frac{2\pi}{|{\bf q}_\parallel|}.
\end{equation}  

Now we consider two charge distributions confined in a slab with finite thickness. We want to treat the real system, which is actually 3D, using an effective 2D description. We do this by depicting the charge distributions as lines of charge (\cref{fig:InteractionSketch}(b)). In other words, we assume that the charge densities are delta functions in-plane and have a certain distribution out-of-plane. The simplest approximation for the out-of-plane distribution is a step function of thickness $d$.  This translates to $\rho_i({\bf r}_\parallel,z)=\frac{q_i\delta({\bf r}_\parallel-{\bf r}_{i,\parallel})}{d}\theta(\frac{d}{2}-|z-z_0|)$, with $z_0$ the center of the material in the perpendicular direction, which leads to an interaction energy of the form (see \cref{appendix:bareQ2D}):
\begin{equation}
\label{eq:bareVQ2D}
V_{Q2D}({\bf q}_\parallel) = \frac{4\pi q_1q_2}{d|{\bf q}_\parallel|^2} \left[1-\frac{2}{|{\bf q}_\parallel|d}e^{-|{\bf q}_\parallel|d/2}\sinh\left(\frac{|{\bf q}_\parallel|d}{2}\right)\right].
\end{equation}
It is instructive to note that in the limit of $q_\parallel d \ll 1$ we recover the 2D potential, while for $q_\parallel d \gg 1$ we get the Coulomb potential for a 3D system (calculated in-plane):
\begin{equation}
V_{Q2D}({\bf q}_\parallel) = \begin{cases}
\frac{2\pi q_1q_2}{|{\bf q}_\parallel|} & q_\parallel d \ll 1 \\
\frac{4\pi q_1q_2}{|{\bf q}_\parallel|^2} & q_\parallel d \gg 1
\end{cases}.
\end{equation}
%
%
%

%So far we have considered the interaction between the charges as if they were alone in the material, but obviously this is not the case in real system and their %expressions have to be modified including the dielectric function. In the next section we show how this can be done.

\section{Screened Interaction}
The (inverse) microscopic dielectric function gives the total potential to first order in the applied external potential,
\begin{equation}
V_{tot}({\bf r},\omega) = \int d{\bf r}' \epsilon^{-1}({\bf r},{\bf r}',\omega)V_{ext}({\bf r}',\omega),
\end{equation}
here a harmonic time-dependence of the fields has been assumed. In standard ab-initio calculations for 3D periodic systems, the dielectric matrix is calculated within the random phase approximation (RPA), which in plane waves representation takes the form: 
\begin{equation}
\label{eq:RPA}
\epsilon_{{\bf G G}'}({\bf q},\omega) = \delta_{{\bf G G}'}-v({\bf q}+{\bf G})\chi^0_{{\bf G G}'}({\bf q},\omega),
\end{equation}
with $v({\bf q}+{\bf G})$ the Fourier transform of the Coulomb Potential and $\chi^0$ the non-interacting response function. 
For a 3D periodic system, the total potential resulting from a plane wave external potential $V_0 e^{i({\bf q\cdot r}-\omega t)}$ has the form 
\begin{equation}
\label{eq:totpotperiodic}
V_{tot}({\bf r},t)=\tilde{V}_{\bf q}({\bf r},\omega)e^{i({\bf q\cdot r}-\omega t)},
\end{equation}
where $\tilde{V}_{\bf q}({\bf r},\omega)$ is a lattice periodic function. Since usually we are interested in macroscopic fields, we define the 3D                                                                                                                                                                                                                                                                                                                                                                                                                                                                                                                                                                                                                                        macroscopic dielectric function as
\begin{equation}
\label{eq:LF3D}
\frac{1}{\epsilon_M({\bf q},\omega)}\equiv \frac{\left\langle\tilde{V}_{\bf q}({\bf r},\omega)\right\rangle_{\Omega}}{V_0}=\epsilon^{-1}_{00}({\bf q},\omega),
\end{equation}
where $\langle ... \rangle_{\Omega}$ denotes a spatial average over a unit cell. Note that in general $\epsilon_M({\bf q},\omega)\neq\epsilon_{00}({\bf q},\omega)$ due to local field effects\cite{Adler1962}.

\subsection{Macroscopic Dielectric Function for 2D Semiconductors}
When \cref{eq:LF3D} is applied to an ab-initio calculation describing a 2D material as an infinite set of parallel sheets separated by a vacuum region of thickness $L$, $\epsilon_M({\bf q},\omega)=1+\mathcal{O}(1/L)$\cite{Falco2013}. This is a consequence of an averaging region much larger than the effective extension of the electron density around the material. The standard definition in \cref{eq:LF3D} becomes meaningless in this case, which is the reason why relatively different values for $\epsilon_M$ have been reported for monolayer MoS$_2$ in the recent literature\cite{Cheiwchanchamnangij2012,Ramasubramaniam2012,Molina2013}. Therefore the definition of the macroscopic dielectric function has to be revised accounting for the finite thickness. From the first equality in \cref{eq:LF3D}, it is natural to substitute an average along the entire unit cell in the out of plane direction with an average over a confined region describing the actual extension of the electronic density. In practice, we average the in-plane coordinates $({\bf r}_{\parallel})$ over the unit cell area $A$ and the out-of-plane coordinate from $z_0-d/2$ to $z_0+d/2$, where $z_0$ denotes the center of the material and $d$ its width.  The macroscopic dielectric function then becomes:
\begin{equation}
\label{eq:dfQ2D}
\begin{split}
\frac{1}{\epsilon_{Q2D}(\textbf{q}_\parallel,\omega)}&\equiv \frac{\left\langle\tilde{V}_{\bf q}({\bf r},\omega)\right\rangle_{A,d}}{V_0} \\
&=\frac{2}{d}\sum_{G_\perp}e^{iG_\perp z_0}\frac{\sin(G_\perp d/2)}{G_\perp}\epsilon^{-1}_{G_\perp~0}(\textbf{q}_\parallel,\omega),
\end{split}
\end{equation}
with $\epsilon^{-1}_{{\bf G G}'}({\bf q}_\parallel,\omega)$ calculated from $\chi^0_{{\bf G G}'}({\bf q}_\parallel,\omega)$ according to the RPA expression in \cref{eq:RPA}. We stress that it is essential to use a truncated Coulomb potential in \cref{eq:RPA} in order to decouple the layers in neighboring supercells\cite{Falco2013}. Note that we used the label Q2D since this definition of macroscopic dielectric function is consistent with the Q2D picture, as we show later on. As a rule of thumb we choose $d$ to be the distance between the layers in the bulk form, but the results for excitons are not very sensitive to this choice as we shown in the next session.

\begin{figure}[t]
\centering  
\includegraphics[width=0.45\textwidth]{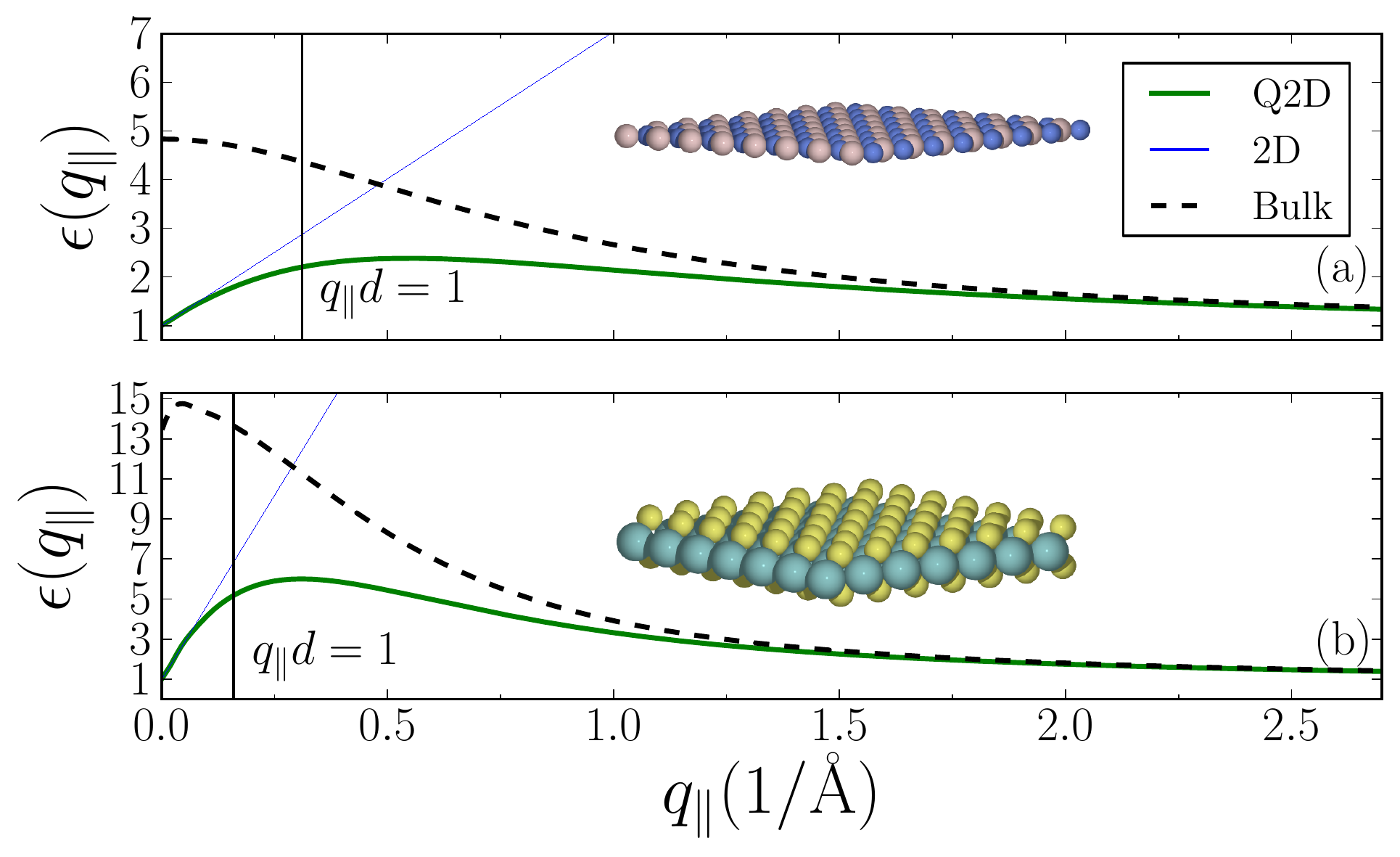}
\caption{Macroscopic dielectric functions for (a) hBN and (b) MoS$_2$. The bulk(black), along with the Q2D (green) and 2D (blue) static dielectric functions are illustrated, the latter corresponding to the linear fits in the small $q_\parallel$ region. For this calculations the $q_\parallel$ values are taken along the $\Gamma-K$ direction, but the homogeneity of the materials has been numerically verified. The parameters used in the linear response ab-initio calculation are discussed in \cref{sec:single_layers}.}
\label{fig:epsM}
\end{figure}

The $q$ dependence of the static dielectric function is illustrated in \cref{fig:epsM} for the case of monolayer hBN and MoS$_2$.
Without loss of generality, the $q_\parallel$ values reported in the plot are chosen to be along the $\Gamma-K$ direction. Indeed, further numerical tests show that the dielectric function is homogeneous, i.e. it is not significantly affected by different direction choices.  
In the low $q_\parallel$ regime the dielectric function approaches one as expected for 2D materials\cite{Falco2013}.
We mention in passing that the dielectric functions of a large collection of 2D materials is available in the Computational Materials Repository\cite{database2015}, see Refs. \onlinecite{Filip2015} and \onlinecite{Andersen2015} .
 
In the plots we also show the linear fit relevant for small $q_\parallel$ as well as the bulk dielectric function. We see that for $q_\parallel d\ll 1$ a linear $\epsilon$ is a viable approximation and we are in a 2D regime. In particular the 2D linear polarizability $\alpha$ can be calculated from the slope of the linear fit. On the other hand, when $q_\parallel d\gg 1$, the bulk behavior of the dielectric function is recovered.

\begin{figure*}[t]
\centering  
\includegraphics[width=1\textwidth]{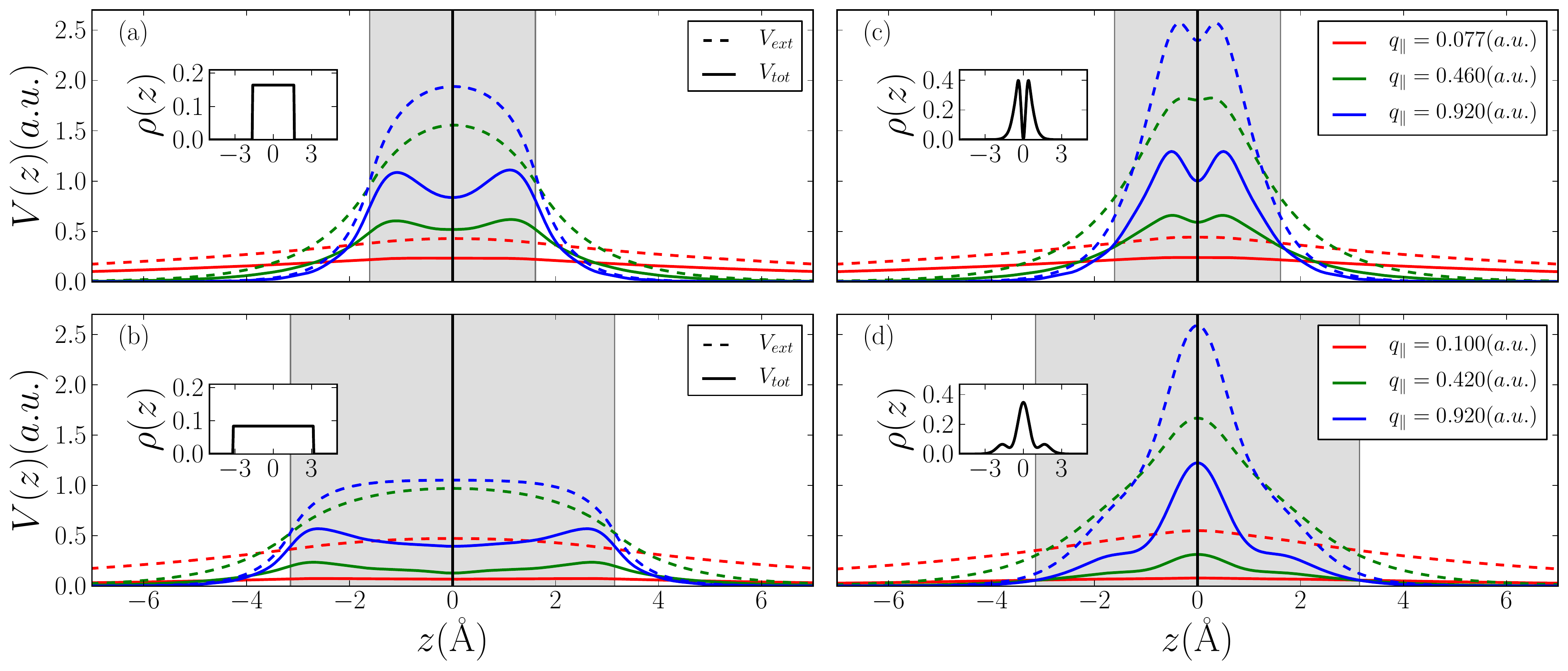}
\caption{$z$-dependence of the total potentials (solid lines) coming from external perturbations (dashed lines) at different in-plane wave-vectors in the case of hBN ((a) and (c)) and MoS$_2$ ((b) and (d)). Left panels: the external perturbation is generated by a step function density distribution (insets). Right panels: the external perturbation is generated by the actual hole out-of-plane density distribution (insets), which is calculated as $\rho_h(z)=\int_A d{\bf r}_\parallel |\psi_{vK}({\bf r}_\parallel,z)|^2$ with $v$ indicating the valence band and K the high symmetry point of the first Brillouin zone. In all cases the density distributions are normalized to 1.}
\label{fig:linear_response}
\end{figure*}

\subsection{Screened Potential in Reciprocal Space}
To account for the screening in the charge-charge interaction we modify \cref{eq:bareVgeneral} introducing the dielectric function:
\begin{equation}
W_{12} = \int_V d{\bf r}d{\bf r}'d{\bf r}''\frac{\rho_1({\bf r})\epsilon^{-1}({\bf r},{\bf r}'')\rho_2({\bf r}')}{|{\bf r''}-{\bf r}'|}.
\end{equation}
In the following, we specialize to the case of electron-hole interaction. Assuming an in-plane delta function distribution and an unspecified $z$-dependence for the charge densities we can easily work out an expression for the screened interaction potential in reciprocal space:
\begin{equation}
\label{eq:Wzgen}
W({\bf q}_\parallel) = \frac{1}{|{\bf q}_\parallel|^2}\int_{-\infty}^{\infty}dzdz' \rho_e(z,{\bf q}_\parallel)\epsilon^{-1}_{00}(z,z',{\bf q}_\parallel)\phi_h(z',{\bf q}_\parallel).
\end{equation}
Here $\rho_e(z,{\bf q}_\parallel)$ is the out-of-plane density distribution for the electron and $\phi_h(z,{\bf q}_\parallel)$ is the out-of-plane potential generated by the hole. For details on how this potential is calculated see \cref{appendix:Poisson}. To study excitons in hBN and MoS$_2$, we take the out-of-plane electron and hole distributions to be $\rho_{e,h}(z)=\int_A d{\bf r}_\parallel |\psi_{c,v~ K}({\bf r}_\parallel,z)|^2$, with $c$ and $v$ the  conduction and valence band indices respectively and $K$ the high symmetry point of the first Brillouin zone, since for both materials that is where the lowest bound exciton is localized \cite{Huser2013,Wang2012}. Furthermore, in \cref{eq:Wzgen} we have introduced a mixed representation for the dielectric function, specifically:
\begin{equation}
\epsilon^{-1}_{00}(z,z',{\bf q}_\parallel)=\frac{1}{L} \sum_{G_\perp G_\perp '}e^{iG_\perp z}\epsilon^{-1}_{0G_\perp 0G_\perp '}({\bf q}_\parallel)e^{-iG_\perp'z'}.
\end{equation}
Note that taking $G_\parallel=G_\parallel'=0$ corresponds to an in-plane macroscopic dielectric function, which also accounts for local field effects.

\begin{figure}[b]
\centering  
\includegraphics[width=0.45\textwidth]{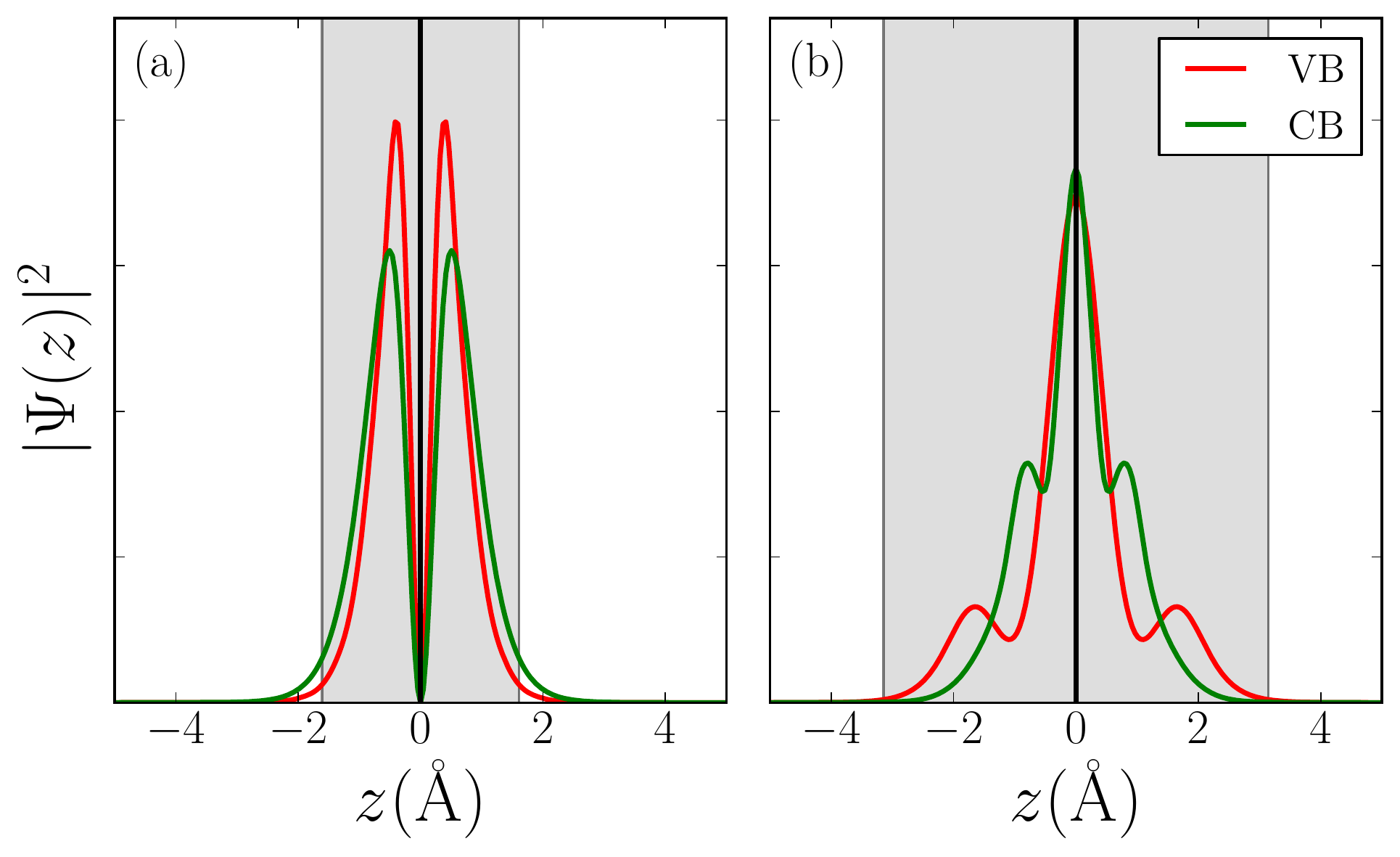}
\caption{Valence (red) and conduction (green) band densities for (a) hBN and (b) MoS$_2$ calculated at the K point.}
\label{fig:WFS}
\end{figure}

To illustrate the effect of screening, \cref{fig:linear_response} shows how a potential generated by either the step function density distribution or the actual hole density distribution is screened by hBN and MoS2. In all cases the density distribution is normalized to 1. The possibility of using either the actual electron/hole out-of-plane density distribution (\cref{fig:WFS}) or a simply step-function gives us two different approximations to calculate the screened interaction within the Q2D picture. 

In the case of step-function density distributions, we can find an analytic expression for the screened potential in \cref{eq:Wzgen}, if we make a further approximation. Indeed, if instead of considering the full $z$-dependence of $\phi_h(z,{\bf q}_\parallel)$ we take its average value within a region of thickness $d$ around the layer, and then screen the resulting constant potential by the full $z$-dependent dielectric function, the general expression \cref{eq:Wzgen} reduces to (see \cref{appendix:screenedQ2D}):
\begin{equation}
\label{eq:WQ2D_q}
\begin{split}
W_{Q2D}({\bf q}_\parallel) =& -\frac{4\pi}{d|{\bf q}_\parallel|^2}\epsilon_{Q2D}^{-1}({\bf q}_\parallel) \times\\
&\left[1-\frac{2}{|{\bf q}_\parallel|d}e^{-|{\bf q}_\parallel|d/2}\sinh\left(\frac{|{\bf q}_\parallel|d}{2}\right)\right]\\
&=\epsilon_{Q2D}^{-1}({\bf q}_\parallel)V_{Q2D}({\bf q}_\parallel),
\end{split}
\end{equation}
where $\epsilon_{Q2D}^{-1}({\bf q}_\parallel)$ is the static version of the macroscopic dielectric function defined in \cref{eq:dfQ2D}.
We thus see that $\epsilon_{Q2D}$ is the natural dielectric function to be used in the quasi-2D picture.

\begin{figure}[t]
\centering  
\includegraphics[width=0.45\textwidth]{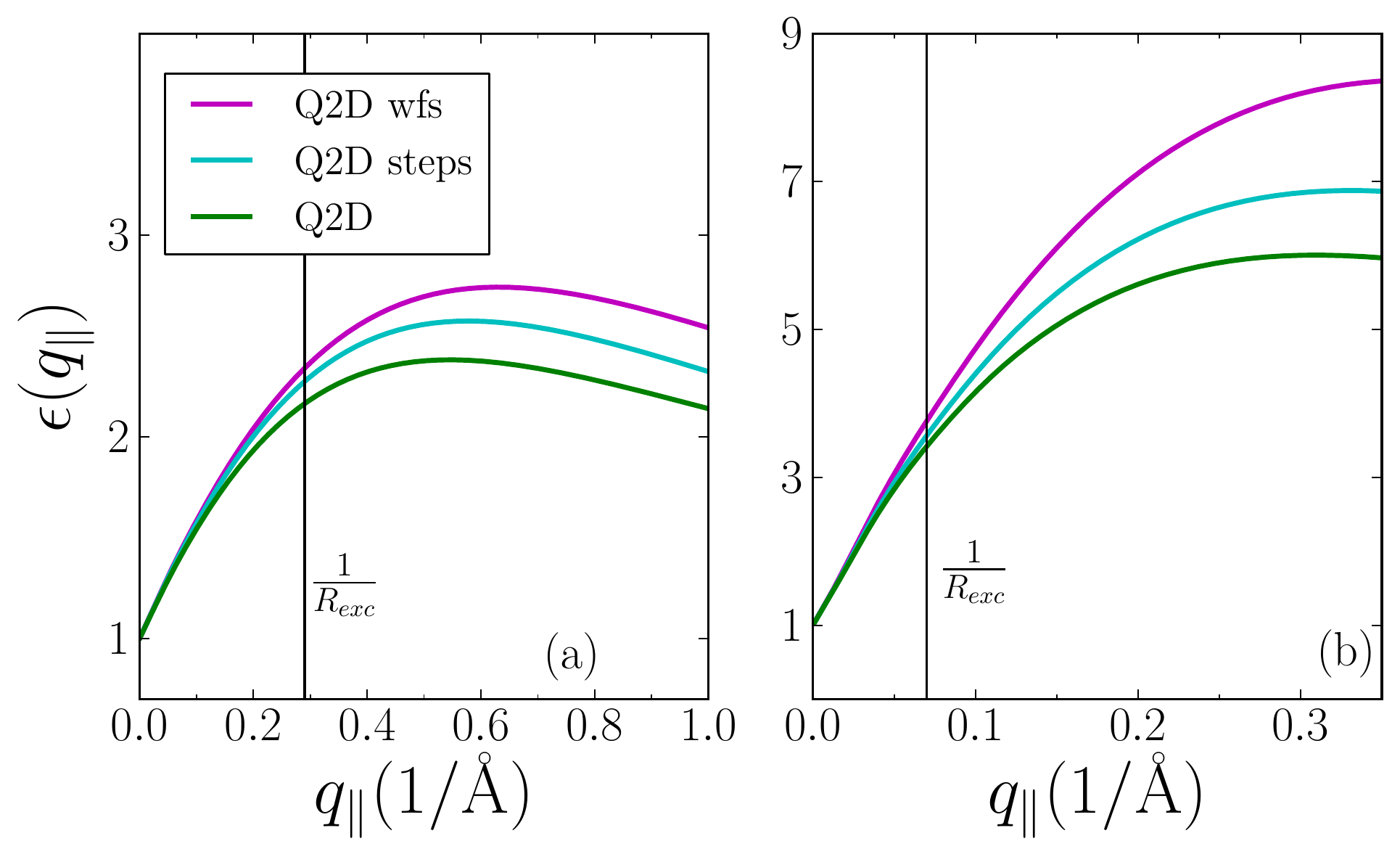}
\caption{Macroscopic dielectric functions for (a) hBN and (b) MoS$_2$. The different dielectric functions are calculated with the three different approaches explained in the text: dielectric function from actual electron and hole distributions (magenta), dielectric function from step function distributions (cyan) and Q2D dielectric function (green). The vertical line represent the radius of the exciton in reciprocal space.}
\label{fig:eps_comparison}
\end{figure}

For each of the two different Q2D models for the screened electron-hole interaction, we can associate a Q2D dielectric function defined as the ratio between the bare and the screened potential:
\begin{equation}
\label{eq:epsgenQ2D}
\epsilon_{Q2D}^{\gamma}({\bm q}_\parallel) = \frac{ \langle \rho_e^\gamma({\bf q}_\parallel)| \phi_h^\gamma({\bf q}_\parallel)\rangle}{ \langle \rho_e^\gamma({\bf q}_\parallel)|\epsilon^{-1}_{00}(\hat{z},\hat{z}', {\bf q}_\parallel)| \phi_h^\gamma({\bf q}_\parallel)\rangle},
\end{equation}
where for simplicity we have used a bracket notation for the integration over $z$ and $\gamma=$~steps,wfs indicates whether the potentials are calculated from step functions or actual electron and hole density distributions. \Cref{fig:eps_comparison} shows a comparison of the two dielectric functions thus obtained together with $\epsilon_{Q2D}$ from \cref{eq:dfQ2D} for hBN and MoS$_2$. 
Clearly the curves perfectly agree in the low $q_\parallel$ regime, while deviations appear for higher values. This observation is consistent with the fact that for small wave vectors the total potentials are flat, and therefore well approximated by the Q2D average value (see \cref{fig:linear_response}).  As we show later, the relevant $q_\parallel$ region for the screening is the one below the the black vertical line representing the inverse exciton radius, calculated from the \emph{ab-initio} Bethe-Salpeter Equation (BSE) (see \cref{sec:single_layers}). Therefore the three different Q2D approaches can be considered equivalent when dealing with excitons in these monolayer materials.

\subsection{Screened potential in real space}
To obtain the form of the screened potentials in real space we Fourier transform \cref{eq:Wzgen}:
\begin{equation}
\label{eq:effpotQ2D}
\begin{split}
W_{Q2D}({\bf r}_\parallel)=&-\frac{2}{d}\int_0^\infty dq \frac{J_0(q|{\bf r}_\parallel|)}{q}\epsilon_{Q2D}^{-1}(q) \times \\
&\left[1-\frac{2}{qd}e^{-qd/2}\sinh\left(\frac{qd}{2}\right)\right],
\end{split}
\end{equation}
where $J_0(x)$ is the zeroth order Bessel function and where we used that the dielectric function is isotropic. This is the quasi-2D potential which can be compared to its strict 2D counterpart defined in \cref{eq:2Dpotq}\cite{Cudazzo2011}:
\begin{equation}
\label{eq:litpot}
W_{2D}({\bf r}_\parallel) = \frac{1}{4\alpha}\left[H_0(x)-N_0(x)\right]_{x=r/2\pi\alpha},
\end{equation}
where $H_0(x)$ and $N_0(x)$ are the Struve and Neumann functions respectively. We stress here that the parameter $\alpha$ can be estimated from the slope of the fit in \cref{fig:epsM}. We note that while this procedure of calculating the 2D polarizability differs from the standard one, it is equivalent. In the case of MoS$_2$, for example, we obtain a value of $5.9$\AA ~which agrees well with the value of $6.6$\AA ~obtained in the literature \cite{Hybertsen2013}.

\begin{figure}[t]
\centering  
\includegraphics[width=0.45\textwidth]{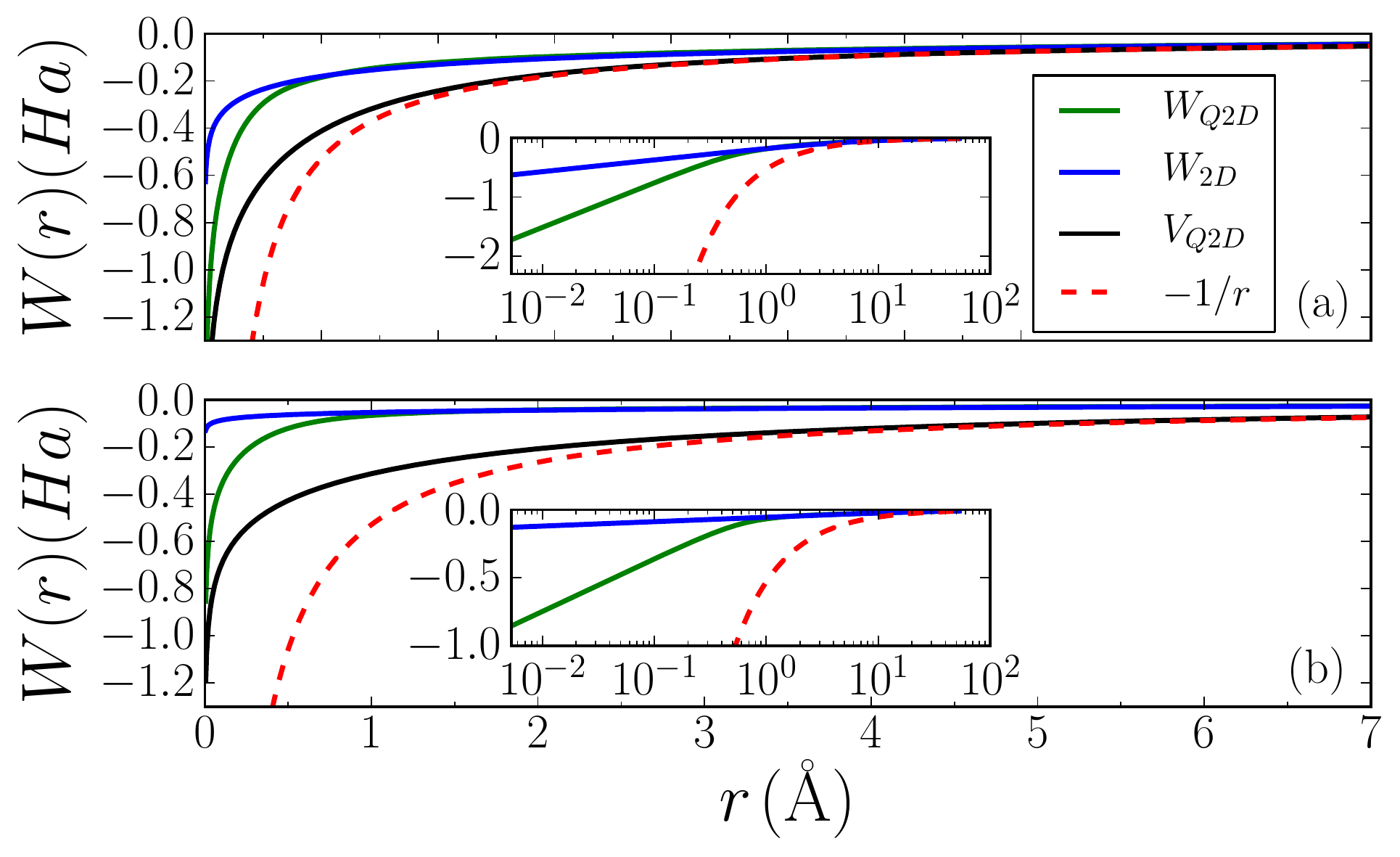}
\caption{Screened Q2D and 2D potentials for (a) hBN and (b) MoS$_2$. The potentials are calculated numerically starting from the macroscopic dielectric functions in \cref{fig:epsM} and using \cref{eq:effpotQ2D} and \cref{eq:litpot} respectively. The bare Q2D curves are calculated using the first equation but neglecting the screening.}
\label{fig:Potentials}
\end{figure}

In \cref{fig:Potentials} we report the numerical results for different potentials: the bare Q2D (black) obtained from  \cref{eq:effpotQ2D} setting $\epsilon_{Q2D}$ to $1$, the screened Q2D (green) obtained from the same equation but including the screening as $\epsilon_{Q2D}$ and the screened 2D (blue) calculated from \cref{eq:litpot}.  The results are shown for both hBN and MoS$_2$.
%
%
%
%Looking at at the bare $Living in a Q2D world affect respect to 1/r
%adding screening changes the interaction even more
%linear and full differ from each other
%Observe that in the long range the potential approach 1/r meaning that the screening has no effect at such long range

We note that the bare Q2D interaction agrees with $-1/r$ beyond a distance given by the layer thickness $d$. Furthermore we see that increasing the layer thickness (going from hBN to MoS$_2$) reduces the bare Q2D interaction strength as expected from \cref{eq:effpotQ2D}. Including the screening reduces the interaction strength even further. The reduction is most significant when using the linear dielectric function (strict 2D screening) as expected from \cref{fig:epsM} which shows that $\epsilon_{2D}(q)>\epsilon_{Q2D}(q)$ for all $q$. We see that, apart from a significant deviation for electron-hole separation smaller than roughly  $1$\AA~, the 2D and Q2D screened potentials agree and both show a logarithmic dependence for $r\rightarrow 0$. For distances larger than the layer thickness, all the potentials (screened and bare) approach the same value ($-1/r$), meaning that screening is completely absent in the asymptotic limit.

\subsection{Importance of the Thickness Parameter}
\begin{figure}[t]
\centering  
\includegraphics[width=0.45\textwidth]{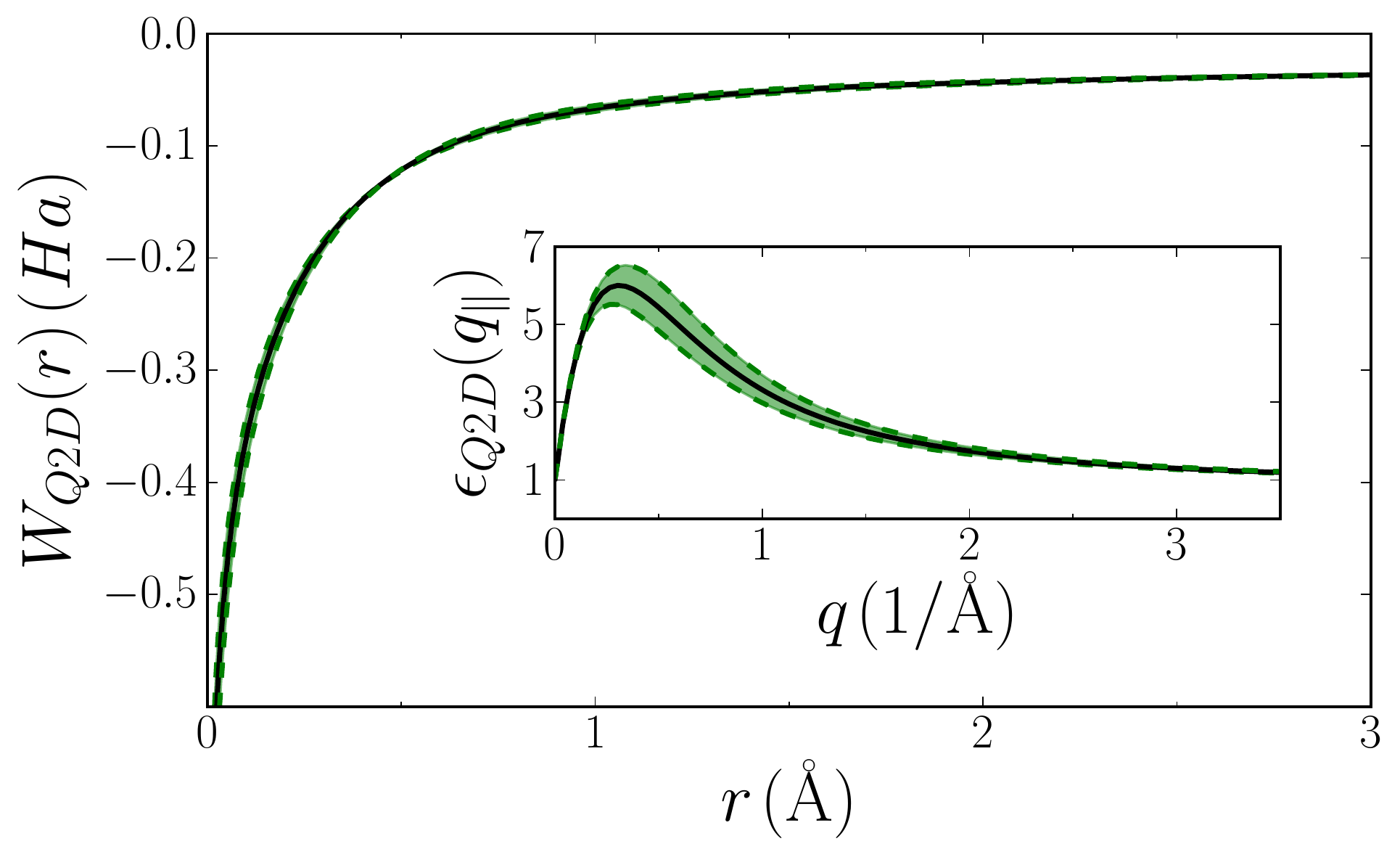}
\caption{Variation of the macroscopic dielectric function and effective potential in MoS$_2$ due to the change in the thickness $d$ of the averaging region in the Q2D model. The continuous black lines are relative to $d=6.29$\AA ~(the interlayer distance in the bulk), while the dashed lines delimiting the shaded region are calculated with a variation of $\pm10\%$ in $d$. }
\label{fig:MoS2_deviation}
\end{figure}

We now return to the problem of choosing the external parameter $d$ entering the Q2D dielectric function. In \cref{fig:MoS2_deviation} we show the Q2D dielectric function and the corresponding potentials when $d$ is varied by $\pm10\%$ around the interlayer distance in bulk MoS$_2$. To the left of the maximum, $\epsilon_{Q2D}$ is insensitive to $d$ since the induced potential is constant over the averaging region. Also in the high $q_\parallel$ limit, $\epsilon_{Q2D}$ is not affected. This is because for these wave vectors the induced potential is in practice negligible. In general, increasing (decreasing) $d$,  decreases (increases) $\epsilon_{Q2D}$ in the large wave vector region. Despite the fact that the variation in the dielectric function is fairly visible for intermediate $q$-values, the screened potential is barely modified. This is because the bare Q2D potential shows an opposite dependence on $d$, such that the product $W_{Q2D}(q)=\epsilon_{Q2D}^{-1}(q)V_{Q2D}(q)$ stays essentially unchanged. In terms of exciton binding energies we have found that a $\pm10\%$ variation in $d$ leads to a correction of less than $0.01eV$. 

\section{Generalized Mott-Wannier Model}
An accurate description of excitonic effects requires the solution of a computationally demanding many-body problem, namely the Bethe-Salpeter equation (BSE)\cite{Strinati1984,Onida2002}. However, it is well known for 3D systems that a satisfying qualitative description can be obtained modelling the exciton as a hydrogenic atom constituted by an excited electron-hole pair interacting via a statically screened Coulomb interaction. In this section we generalize such a model to the Q2D case.

The Bethe-Salpeter two particle Hamiltonian for a 2D periodic system is given by:
\begin{equation}
\begin{split}
H^{2P}_{\substack{n_1n_2{\bf k}_1\\ n_3n_4{\bf k}_2}}({\bf q}_\parallel)=&(\epsilon_{n_2{\bf k}_1+{\bf q}_\parallel}-\epsilon_{n_1{\bf k}_1})\delta_{n_1n_3}\delta_{n_2n_4}\delta_{{\bf k}_1{\bf k}_2}+ \\
&-(f_{n_2{\bf k}_1+{\bf q}_\parallel}-f_{n_1{\bf k}_1})K_{\substack{n_1n_2{\bf k}_1\\ n_3n_4{\bf k}_2}}({\bf q}_\parallel),
\end{split}
\end{equation}
where $n_i$ are band indices, ${\bf k}_i$ vectors in the first 2D Brillouin zone and ${\bf q}_\parallel$ is the in-plane momentum transfer, or exciton center-of-mass momentum. In the following we specialise to the case of vertical transitions, i.e. ${\bf q}_\parallel= 0$. $K$ is the kernel containing the exchange and the screened direct Coulomb interaction.
This Hamiltonian describes scattering processes between two electron-hole pairs excited by an external perturbation. In general these processes should involve all the occupied and unoccupied states in the spectrum; however, when the conduction and valence bands are well separated from the remaining bands, it is often a good approximation to include only the valence and conduction band states. Together with the Tanm-Dancoff approximation this assumption allows us to express the resonant part of the two-particle Hamiltonian as:
\begin{equation}
H^{2P (res)}_{\substack{vc{\bf k}\\ vc{\bf k}'}}=(\epsilon_{c{\bf k}'}-\epsilon_{v{\bf k}})\delta_{{\bf k}{\bf k}'}-K_{\substack{vc{\bf k}\\ vc{\bf k}'}}.
\end{equation}
The kernel is given by 
\begin{equation}
\label{eq:kernel}
\begin{split}
K_{\substack{vc{\bf k}\\ vc{\bf k}'}}=&-\int_A d{\bf r}d{\bf r}'\psi_{v{\bf k}}({\bf r})\psi_{c{\bf k}}^*({\bf r}')W({\bf r},{\bf r}')\psi_{v{\bf k}'}^*({\bf r})\psi_{c{\bf k}'}({\bf r}')+\\
&2\int_A d{\bf r}d{\bf r}'\psi_{v{\bf k}}({\bf r})\psi_{c{\bf k}}^*({\bf r})v({\bf r},{\bf r}')\psi_{v{\bf k}'}^*({\bf r}')\psi_{c{\bf k}'}({\bf r}')
\end{split}
\end{equation}
where $|\psi_{\alpha {\bf k}}\rangle$, with $\alpha=(v,c)$, represents single particle Bloch states for the valence and conduction band, $W({\bf r},{\bf r}')=\int d{\bf r}''\frac{\epsilon^{-1}({\bf r},{\bf r}'')}{|{\bf r}''-{\bf r}'|}$ is the screened interaction and $v({\bf r},{\bf r}') = \frac{1}{|{\bf r}-{\bf r}'|}$ is the bare Coulomb interaction. The first term on the right hand side of \cref{eq:kernel} is the direct screened electron-hole interaction while the second is the Coulomb exchange. Our full \emph{ab-initio} solution of the BSE shows that the exchange term only slightly decreases the exciton binding energy by 0.08 eV and 0.02 eV for hBN and MoS$_2$, respectively. This amounts to less than $5\%$ of the total binding energy and we therefore neglect the exchange contribution in the rest of the paper.

Throughout the BZ we consider the valence and conduction band wave functions to be plane waves in the in-plane direction and in the out-of-plane direction equal to $\psi_\perp(z)=\left(\int_A d{\bf r}_\parallel |\psi_{\alpha K}({\bf r}_\parallel,z)|^2\right)^{1/2}$ up to a normalization factor and with $\alpha=v,c$. 
With this approximation and proceeding as for \cref{eq:Wzgen}, the interaction matrix becomes: 

\begin{equation}
\label{eq:Ukkfinal}
K_{\substack{vc{\bf k}\\ vc{\bf k}'}} = \frac{1}{A}W(|{\bf k}-{\bf k}'|),
\end{equation}
where $W(|{\bf k}|)$ is the screened potential in \cref{eq:Wzgen} which can be evaluated in the various ways described in the previous section, depending on the level of approximation.

Completely analogue to the 3D case we can introduce the envelope function $F({\bf r}_{\parallel})$, defined as $F({\bf r}_{\parallel})=\sum_{k}e^{-i{\bf k}{\bf r}_{\parallel}}A({\bf k})$, with $A({\bf k})$ excitonic weights in reciprocal space, and arrive at an eigenvalue problem of a 2D hydrogenic atom,
\begin{equation}
\label{eq:MWHamiltonian}
\left[-\frac{\nabla_{2D}^2}{2\mu_{ex}}+W({\bf r}_\parallel)\right]F({\bf r}_\parallel)=E_bF({\bf r}_\parallel),
\end{equation}
where $\mu_{ex}$ is the exciton effective mass, calculated from the hole and electron masses according to $\mu_{ex}^{-1}=m_{e}^{-1}+m_{h}^{-1}$.

\section{Exciton binding Energies of isolated monolayers}
\label{sec:single_layers}
In this section we investigate the performance of the Mott-Wannier model in \cref{eq:MWHamiltonian} for the calculation of binding energies of the lowest bound exciton in hBN and MoS$_2$. 

\subsection{Ab-initio Calculation Details}
In order to solve \cref{eq:MWHamiltonian} with either the Q2D or 2D potentials, we need to calculate the dielectric matrix. We describe the two materials with a supercell technique and we optimize the structure using the LDA exchange-correlation potential; geometrical details are provided in \cref{tab:geom}.
\begin{table}[t]
\label{tab:geom}
\caption{Geometry and effective masses.}
\resizebox{0.3\textwidth}{!}{
\begin{tabular}{ccccc}
\hline
Material & $a$ (\AA) & $L$(\AA) & $d$ (\AA) & $\mu^{ex}$ (a.u.) \\ \hline
MoS$_2$ & 3.20 & 23.0 & 6.29 & 0.27 \\ 
hBN & 2.50 & 23.0 & 3.22 & 0.37  \\ 
 \hline
\end{tabular}}
\end{table}
To calculate the non-interacting response function we use $150 eV$ cut-off energy for the reciprocal lattice vectors $\bf G$ and $\bf G'$ in order to account for local field effects. We construct $\chi^0$ from local density approximation (LDA) wave functions and energies, and we then get the dielectric matrix using a truncated Coulomb potential in order to avoid interaction between supercells \cite{Rozzi2006}. The dielectric matrix is calculated on a $60\times60$ k-points grid. Since it turns out that the exciton binding energy is sensitive to the low wave-vector behavior of the screening, we use an expansion of the density response function $\chi^0$ around $q_\parallel=0$ in order to calculate the dielectric matrix in the small $q_\parallel$ limit. All calculations are performed with the GPAW code\cite{Enkovaara2010,GPAW_web}, which is based on the projector augmented wave method. Details on the implementation of the linear response code can be found in Ref.\onlinecite{Yan2012}. We mention that the dielectric functions of more than fifty 2D materials calculated in this fashion are available in the CMR\cite{database2015}. The exciton masses as computed from the LDA band structure are given in \cref{tab:geom}. 

\begin{table}[b]
\label{tab:EbResults}
\caption{Numerical values for energy of the lowest bound excitonic state at the direct gap. Both the BSE and the models are based on LDA ab-initio calculations. The exchange contribution is not included.}
\resizebox{0.45\textwidth}{!}{
\begin{tabular}{lccccc}
\hline
& $\bf E_b^{BSE}(eV)$ & $\bf E_b^{Q2D}(eV)$ & $\bf E_b^{2D} (eV)$ & $\bf E_b^{steps}(eV)$  & $\bf E_b^{wfs}(eV)$ \\ \hline
$\bf hBN$ & 2.05 & 2.35 & 2.34 & 2.23 & 2.29  \\ 
$\bf MoS2$ & 0.43 & 0.61 & 0.60 &0.57 & 0.59\\ 

\hline
\end{tabular}}
\end{table}

\begin{figure}[t]
\centering  
\includegraphics[width=0.45\textwidth]{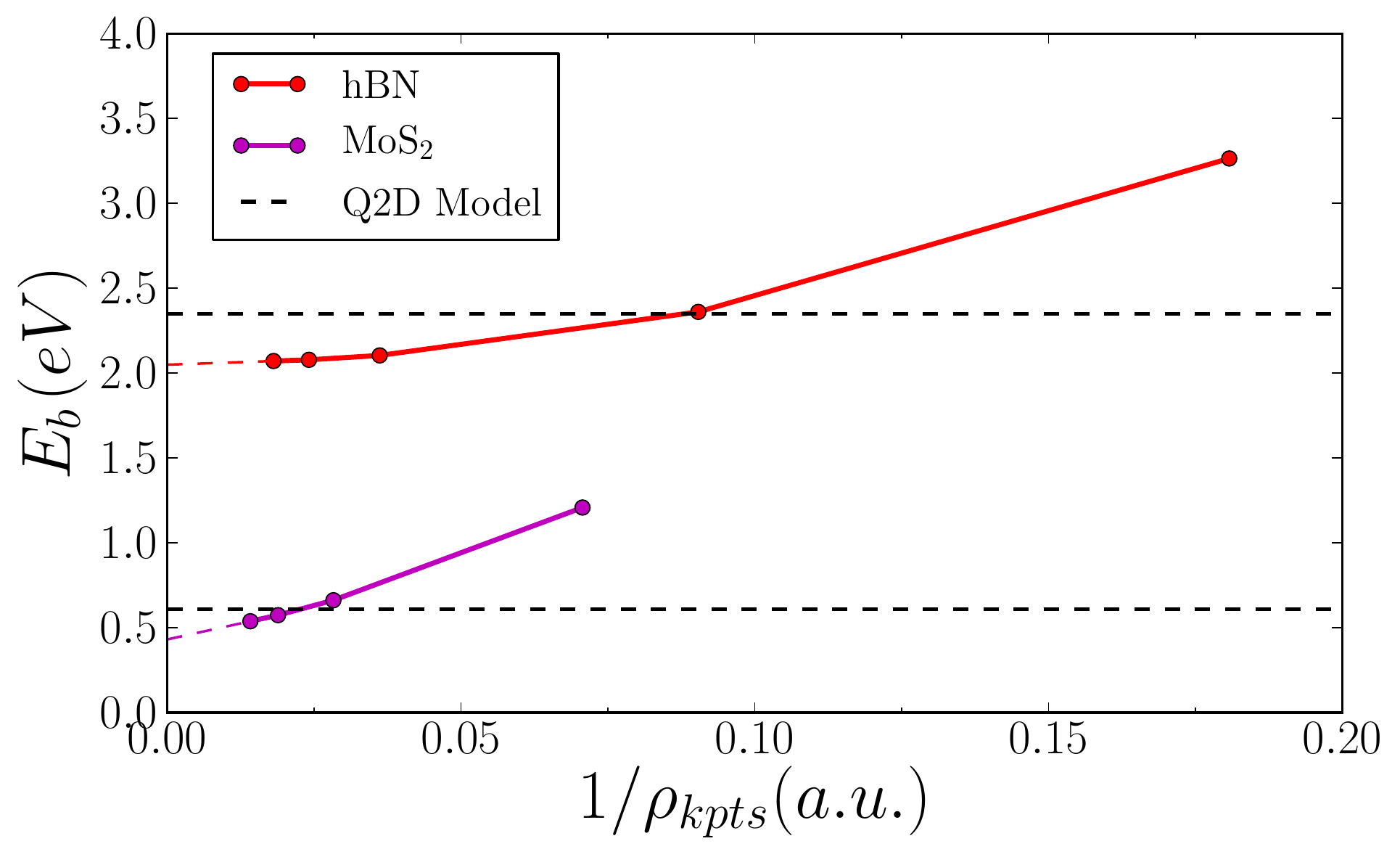}
\caption{Convergence plot for the binding energy obtained from the BSE solution against the k-points density. Extrapolation to infinite k-point sampling is shown. In the BSE the exchange contribution is left out. The horizontal dashed lines show the results given by the Q2D model.} 
\label{fig:BSEconv}
\end{figure}

To obtain the lowest bound exciton we numerically solve the Mott-Wannier equation on a logarithmic grid. With this method we are able to converge the lowest eigenvalue with a precision of $0.002$eV. For benchmark, we perform BSE calculations using the GPAW code. For the screening of the electron-hole interaction we use the static dielectric function evaluated with the same parameters employed in the linear response calculation. The particle-hole states of the BSE Hamiltonian are constructed from a single LDA valence and conduction band. To compare directly to our model, all the BSE calculations are performed neglecting the exchange part of the kernel. We stress that the binding energy of the first exciton changes by less than $0.01 eV$ if the BSE Hamiltonian is constructed from the four highest and four lowest conduction bands. As reported previously\cite{Falco2013,Qiu2013}, BSE binding energies in 2D materials are extremely sensitive to the $k$-point grid. We therefore perform BSE calculations with up to $60\times 60$ $k$-points, for which we get binding energies of $2.07$eV and $0.54$eV for hBN and MoS$_2$ respectively. Furthermore, assuming a linear dependence of the binding energy with respect to $1/\rho_{kpts}$, we extrapolate the results to infinite $k$-points sampling (see \cref{fig:BSEconv}).
%
%
%
\begin{comment} The linear extrapolation can be justified by means of Q2D BSE calculations, i.e. calculations performed with a model BSE in reciprocal space which uses parabolic bands and a the kernel in \cref{eq:Ukkfinal}. We show the high k-point density limit convergence test made with the Q2D BSE in the inset of \cref{fig:BSEconv}. From here we see that a linear extrapolation perfectly matches the real space result (not affected by convergence issues) and therefore we can safely extend it to the full BSE case.
\end{comment}

\subsection{Results}
The values for binding energy of the lowest bound exciton obtained with the different models for the screened electron-hole interaction along with the extrapolation from the BSE are reported in \cref{tab:EbResults}. 
We first observe that there is practically no difference in the binding energies obtained from the Mott-Wannier model using either the Q2D or 2D screened potentials. Moreover, the result from the Mott-Wannier model(s) are within $0.3eV$ and $0.18eV$ of the BSE result for hBN and MoS$_2$ respectively. We consider this a reasonable agreement given the simplicity of the model.  

In \cref{tab:EbResults} we also report the binding energies obtained when the electron-hole interaction is calculated numerically from \cref{eq:Wzgen} using step-functions and actual electron and hole density distributions. As pointed out in the discussion of \cref{fig:eps_comparison} we expect these two other approaches to give the same description of excitons. Indeed, the binding energies we obtained are in perfect agreement with the Q2D and 2D model results.

\begin{figure*}[t]
\centering  
\includegraphics[width=1\textwidth]{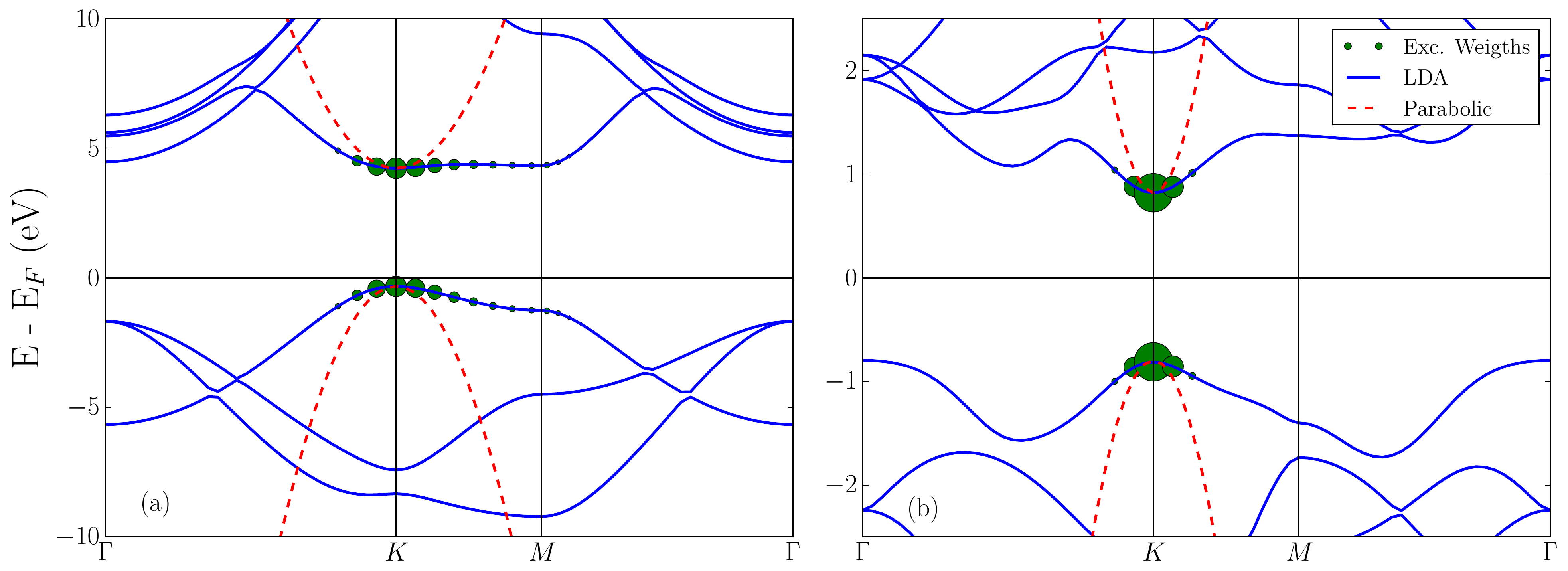}
\caption{LDA band structure and exciton weights for (a) hBN and (b) MoS$_2$. In both materials the exciton is well localized at the K point. The excitonic weights are calculated as the absolute value squared of the eigenvector of the two particle BSE Hamiltonian associated to the lowest bound exciton. In red the parabolic bands used in the Mott-Wannier model. The values for the electron and hole masses are $0.93 a.u.$ and $0.62 a.u.$ for hBN and $0.61 a.u.$ and  $0.49 a.u.$ for MoS$_2$.}
\label{fig:exciton_weigths}
\end{figure*}

The agreement between the Q2D and 2D descriptions can be understood by looking at the $q$-space extension of the lowest bound exciton wave function shown in \cref{fig:exciton_weigths}. 
We see that for both hBN and MoS$_2$ the exciton is confined to a rather narrow region around the K-point. A localization of the exciton in $q$-space means that the relevant contribution to the electron-hole interaction comes from the low wave-vector regime. From the calculated excitonic wavefunctions in real space we obtain inverse exciton radii of $0.29$\AA$^{-1}$ for hBN and $0.07$\AA$^{-1}$ for MoS$_2$. Both of these values are comparable to $1/d$ ($0.31$\AA$^{-1}$ and $0.16$\AA$^{-1}$, respectively). As we have seen previously, in this limit the Q2D screened potential reduces to the strict 2D potential explaining the similarity of the binding energies obtained with the two descriptions. 

\begin{comment}In addition, the localization of the exciton in reciprocal space justifies the use of a parabolic approximation to describe the band structures.\end{comment}

To conclude this section, we notice that in the evaluation of the screened electron-hole interaction, we neglected the in-plane spatial variation of the conduction and valence band wavefunctions. The validity of this approximation can be checked by performing a BSE calculation where the screened interaction is evaluated using a dielectric matrix $\epsilon_{{\bf G}{\bf G}'}^{-1}$ where all matrix elements except for those where ${\bf G}_{\parallel}={\bf G}_{\parallel}'=0$ are set to zero. In other words, we neglect all the in-plane high frequency spatial variations of the wave functions. With this constriction we obtain a binding energy of $2.21 eV$ for hBN and $0.44$ for MoS$_2$. The neglect of in-plane variations of the wave functions is thus responsible for 0.15 eV (hBN) and 0.01 eV (MoS$_2$) of the observed discrepancy between the Mott-Wannier model and the full BSE calculation.  

\section{Excitons in layered structures}
In this section, we show that a linear approximation for the dielectric function breaks down when applied to excitons in multi-layered structures. While it is possible to include the non-linear q-dependence of the dielectric function within a strict 2D model, the Q2D description turns out to be necessary to quantitatively capture screening effects.

\subsection{The Quantum Electrostatic Heterostructure (QEH) Model}
In order to calculate exciton binding energies in a layered structure we first need the dielectric function. This can be obtained using the quantum-electrostatic heterostructure (QEH) model that we introduced recently\cite{Andersen2015}.  In brief, the underlying procedure in the calculation of the dielectric function can be divided in two parts. In the first part the full density response function of each isolated layer, calculated from first principles, is used to obtain the monopole/dipole components of the density response function as well as the spatial profile of the electron densities in the $z$-direction induced by a monopole/dipole field. We refer to these data sets as the dielectric building block of the individual layer. In the second part, the dielectric building blocks are coupled together via the Coulomb interaction to give the dielectric function for the full structure. The dielectric matrix obtained from the QEH model can be used to obtain the electron-hole interaction according to
\begin{equation}
\label{eq:W_q_vdWH}
W(q_\parallel) = \underline{\rho}_{\hspace{0.05cm}e}^{\intercal}(q_\parallel)~\underline{\underline{\epsilon}}^{-1}(q_\parallel)~\underline{\phi}_{\hspace{0.05cm}h}(q_\parallel),
\end{equation}
where $\underline{\rho}_{\hspace{0.05cm}e}$ ($\underline{\phi}_{\hspace{0.05cm}h}$) is the electron density (hole induced potential) vector expressed in a basis set of monopole/dipole densities (potentials). The basis set of induced densities and potentials is also used as (left and right) basis functions for representing $\underline{\underline{\epsilon}}^{-1}$. To be more explicit an arbitrary density vector $\underline{\rho}$ can be represented as $\underline{\rho}^{\intercal}=[\rho_{1M},\rho_{1D},\rho_{2M},\rho_{2D},\cdots,\rho_{nM},\rho_{nD}]$ where $\rho_{i\alpha}$, with $\alpha=M,D$, is the induced monopole/dipole density at the layer $i$. A completely equivalent expression can be formulated for the induced potentials.

It is clear that the equation above is just a simple rewriting of \cref{eq:Wzgen} in terms of a minimal monopole/dipole basis. 
We point out that this formalism takes the finite extension of the layers in the out-of-plane direction into account, and is therefore consistent with the Q2D picture described in the previous sections. In Ref. \onlinecite{Andersen2015} we showed, based on the comparison with full \emph{ab-initio} calculations, that the monopole/dipole basis is sufficient to obtaining an accurate description of the dielectric and plasmonic properties of different layered heterostructures. 

\subsection{Breakdown of the Linear Screening Model}
\begin{figure}[b]
\centering  
\includegraphics[width=0.45\textwidth]{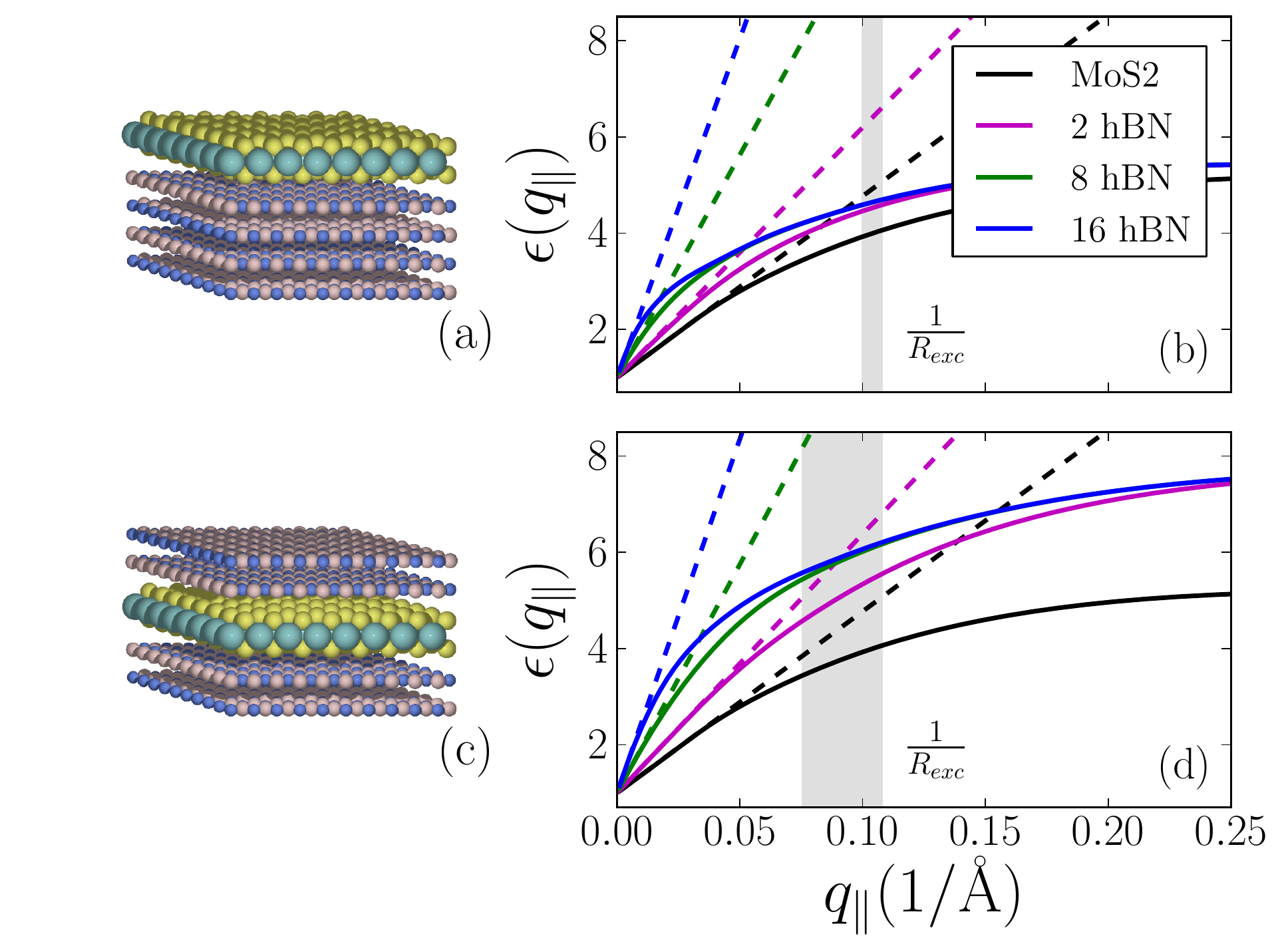}
\caption{Left panels: The on-top (a) and sandwich (c) arrangements of the MoS$_2$/hBN heterostructures. Right panels: effective dielectric function (full line) for the on-top (b)  and sandwich (d) configurations. The linear approximations to the dielectric function is shown by dashed lines. The shaded regions in (b) and (d) represent the range of inverse exciton radii found for the considered structures. The $q$-values below these regions are relevant for screening the electron-hole interaction and for the thicker structures this region extends beyond the linear regime of $\epsilon(q)$.}
\label{fig:VdW_eps}
\end{figure}

As an example we consider two different types of heterostructures. The first, which we refer to as ``on-top'', consists of MoS$_2$ on top of $n$ layers of hBN. The second, which we refer to as ``sandwich'', consists of an MoS$_2$ layer encapsulated in 
$n$ layers of hBN, see \cref{fig:VdW_eps} panel (a) and (c). The interlayer distance between MoS$_2$ and hBN is set to $5.1$\AA. 
In \cref{fig:VdW_eps} panel (b) and (d) we show the dielectric function of the MoS$_2$ layer as well as the linear approximation as a function of the in-plane wave vector for different number of hBN layers. The effective dielectric function of MoS$_2$ in the heterostructure is defined along the lines of \cref{eq:epsgenQ2D}:
\begin{equation}
\label{eq:eps_q_vdWH}
\epsilon(q_\parallel) = \frac{\underline{\rho}_{\hspace{0.05cm}e}^{\intercal}(q_\parallel)~\underline{\phi}_{\hspace{0.05cm}h}(q_\parallel)}{\underline{\rho}_{\hspace{0.05cm}e}^{\intercal}(q_\parallel)~\underline{\underline{\epsilon}}^{-1}(q_\parallel)~\underline{\phi}_{\hspace{0.05cm}h}(q_\parallel)},
\end{equation}
which gives the ratio of the bare to the screened interaction between an electron and a hole in the MoS$_2$ layer.   

From \cref{fig:VdW_eps}, we notice that adding hBN layers to MoS$_2$ changes the shape of the dielectric function introducing a pronounced feature that shifts towards low $q_\parallel$ as the number of hBN layers is increased. This shoulder-like feature can be explained as an interplay between the 3D and 2D screening character. When more hBN layers are added to the heterostructure, the system tends toward a bulk limit, where the dielectric function is larger than 1 for $q_\parallel=0$. However, the heterostructure has a finite thickness $d$ and, as required by the 2D limit, when $q_\parallel\ll 1/d$ the dielectric function is 2D like and becomes 1 for $q_\parallel=0$. This leads to a sharp drop in the dielectric function, which becomes steeper as the thickness of the heterostructure is increased, explaining the appearance of the shoulder-like feature. 
It is clear, from \cref{fig:VdW_eps}, that the main change in the dielectric function is caused by the nearest layers of hBN. Adding more layers only causes a slight variation. Obviously, this is due to the fact that hBN is less effective at screening the electron-hole interaction as the distance from MoS$_2$ is increased. For the same reason, the screening is more pronounced in the sandwich configuration than in on-top configuration.

\begin{figure}[t]
\centering  
\includegraphics[width=0.45\textwidth]{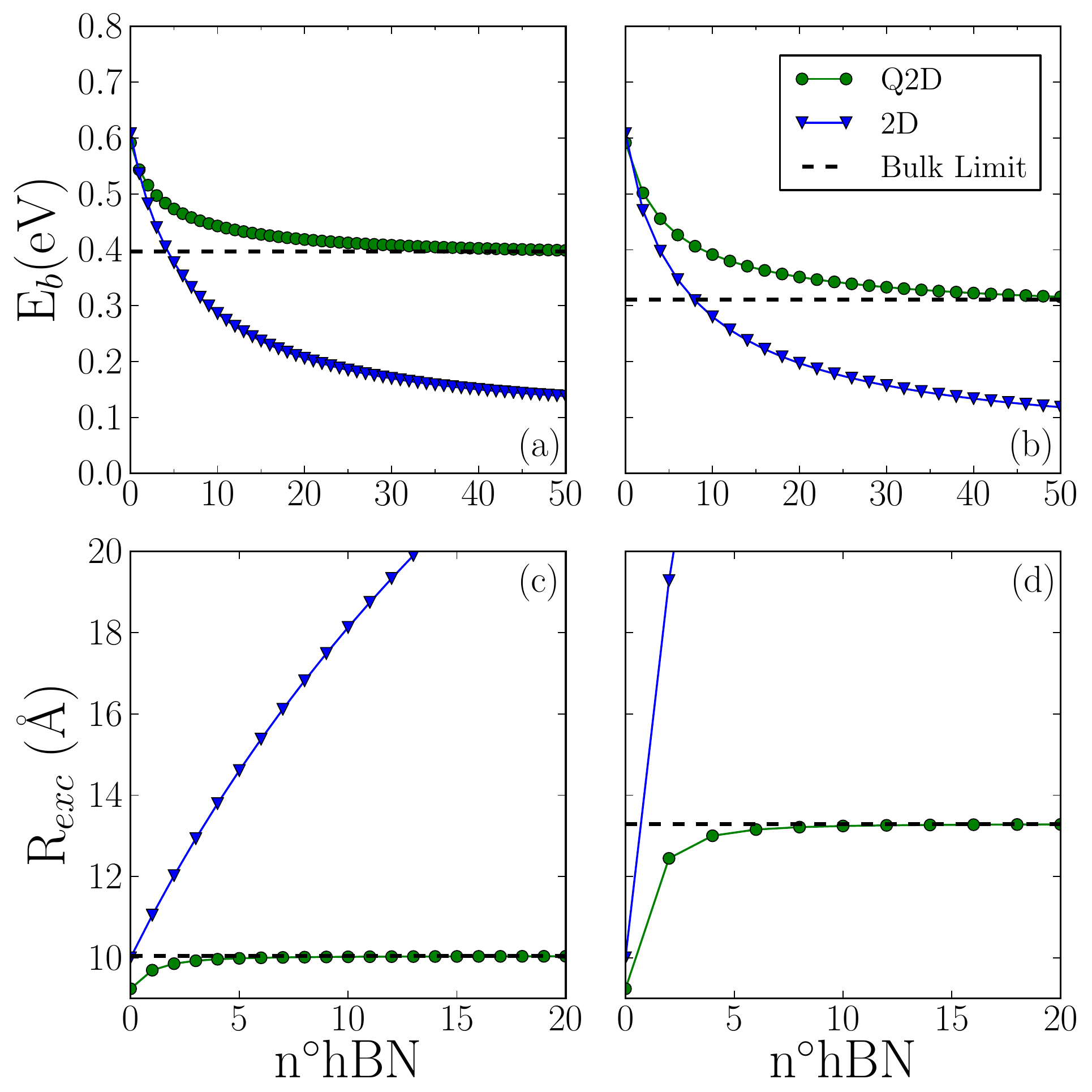}
\caption{Energy and radius of the lowest bound exciton for the (a,c) on-top and (b,d) sandwich configuration as function of the number of hBN layers obtained from the Q2D (green) and 2D (blue) approaches. }
\label{fig:VdW_Eb}
\end{figure}

We then proceed to calculate the binding energy of the lowest bound exciton in the MoS$_2$ layer for the two different configurations using both the full wave vector dependent dielectric function (quasi-2D) and its linear approximation (strict 2D). The results are shown in \cref{fig:VdW_Eb} panel (a) and (b).  When the full dielectric function is used, the binding energy converges towards $0.40eV$  and $0.31eV$ for the on-top and sandwich configurations, respectively. These values represent the bulk limits, i.e. MoS$_2$ on a hBN substrate and MoS$_2$ encapsulated by two seminfinite stacks of hBN. The reduction in binding energy of 0.2 eV for the on-top configuration is in good agreement with the experimentally determined change in exciton energy of WS$_2$ when adsorbed on SiO$_2$\cite{Chernikov2014} (the bulk dielectric constants of SiO$_2$ and hBN are similar). In contrast, the assumption of linear dielectric screening completely fails in estimating the exciton binding energies. Indeed, it quickly diverges from the Q2D results yielding much too small binding energies. This behaviour results from the continuously increasing slope of the dielectric function eventually arriving at a condition of perfect screening (infinite slope).

Figure \ref{fig:VdW_Eb} panels (c) and (d) show the exciton radii obtained from the Q2D and 2D models. Interestingly, for the Q2D model the increase in the exciton radius due to the screening from the hBN is only 10\% and 30\% for the on-top and sandwich configurations, respectively. The range of the inverse exciton radius is indicated by a shaded region in \cref{fig:VdW_eps} panel(c) and (d). As we demonstrated in the previous section, the relevant $q_\parallel$ for the screening lie mainly below the inverse exciton radius. Inspection of \cref{fig:VdW_eps} clearly indicates that in this regime the linear approximation overshoots the full wave vector dependent dielectric function and it gets worse as the number of layers is increased.

\subsection{Limitations of the 2D picture in Layered Structures}
In the previous paragraph we showed that the assumption of linear screening, i.e. \cref{eq:lineps}, breaks down when the screening from the environment is included. It is, of course, possible within the 2D picture to couple a stack of 2D materials, each described by a linear dielectric function, using the QEH model. In this section we explore the validity of such an approach using the Q2D results obtained in the previous section as a reference. 

We model the layered structure as infinitesimally thin planes described by 2D building blocks, as opposed to the Q2D ones used previously, and couple them electrostatically via the QEH. While it is straightforward to define multipole response function and induced density components when a finite thickness is considered, in 2D only the monopole components have an obvious definition. 
Within the 2D picture, the monopole induced density is described by a delta function centered at the layer position $z_i$. The component of the 2D response function of the (isolated) layer may be obtained from the corresponding 2D dielectric function in \cref{eq:lineps}:
\begin{equation}
\begin{split}
\tilde{\chi}_{2D}^{M}(q_\parallel)&=\frac{q_\parallel}{2\pi}\left[\epsilon_{2D}^{-1}(q_\parallel)-1\right]\\
&=-\frac{\alpha q_\parallel^2}{1+2\pi\alpha q_\parallel}.
\end{split}
\end{equation}
For strict 2D layers, the Coulomb interaction between monopole charge densities in layers at $z_i$ and $z_k$ takes the form 
\begin{equation}
V_{iM,kM}(q_\parallel)=\frac{2\pi e^{-q_\parallel|z_i-z_k|}}{q_\parallel},
\end{equation}
which reduces to the standard 2D Coulomb potential in reciprocal space for coupling within the layer.

\begin{figure}[t]
\centering  
\includegraphics[width=0.45\textwidth]{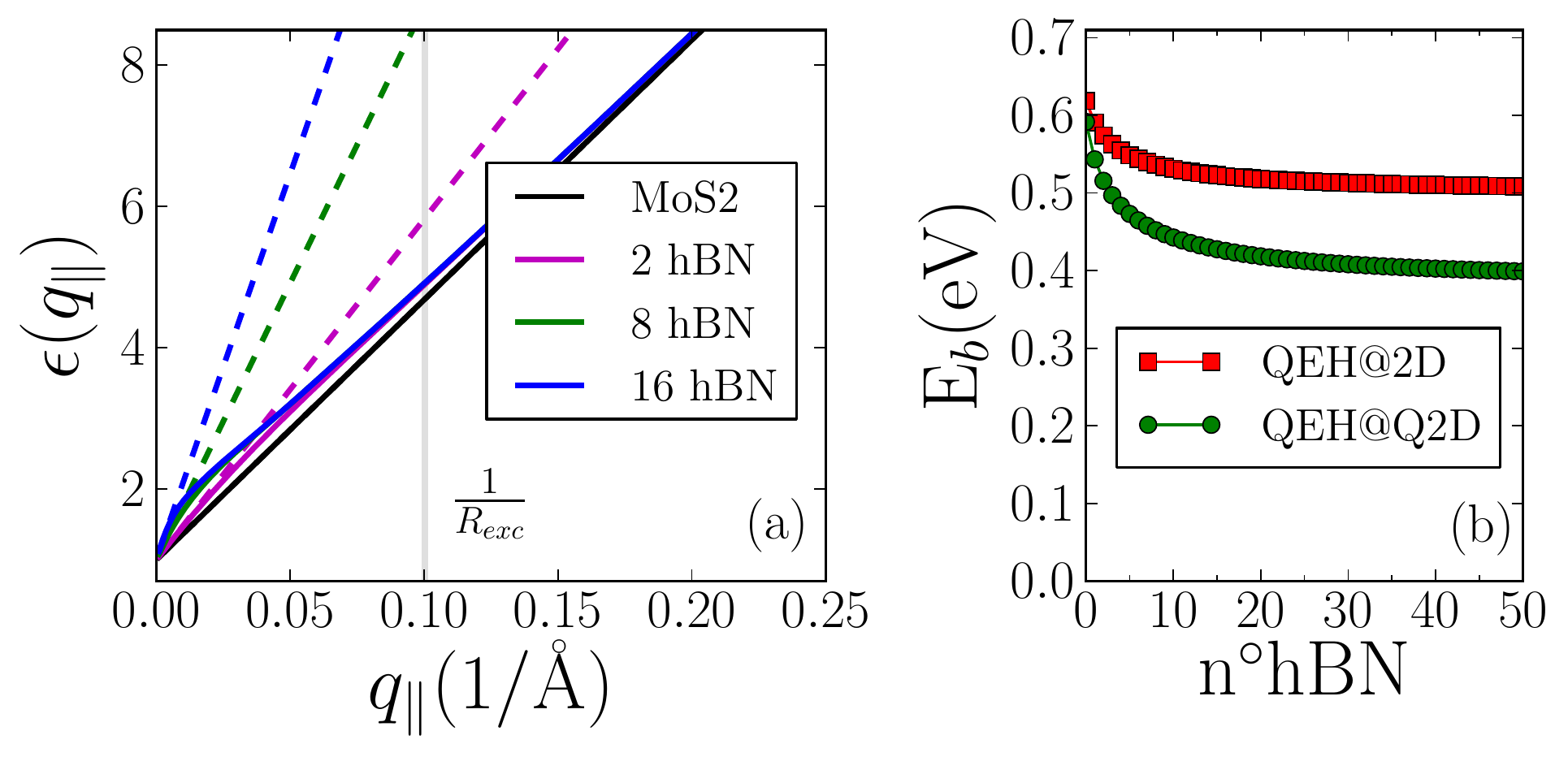}
\caption{(a) Effective dielectric function (full line) and (b) energy  of the lowest bound exciton for the on-top MoS$_2$-hBN configuration, calculated with the QEH model based on a 2D description of the layers. The linear approximations to the dielectric function is shown by dashed lines in panel (a), along with the range of inverse exciton radii found for the considered structures represented by the shaded region.}
\label{fig:VdW_eps_QEH2D}
\end{figure}

To test the QEH with 2D building blocks, we consider the ``on-top'' structure of the previous paragraph and in \cref{fig:VdW_eps_QEH2D} we report the effective dielectric function and energy of the lowest bound exciton as a function of the number of hBN layers.
It is clear that the 2D dielectric function of the supported MoS$_2$ deviates significantly from the Q2D result (see \cref{fig:VdW_eps} panel (b)) for larger $q_\parallel$. However, for smaller $q_\parallel$ the 2D function actually reproduces qualitatively the non-linear structure of the Q2D result. In terms of exciton binding energy, we observe a convergence to a finite value when the number of layers of hBN is increased, but the reduction in binding energy compared to the free-standing layer is $50\%$ smaller than the reduction obtained with the Q2D approach. The underestimation of the screening can be ascribed essentially to two reasons.  First, the potential generated by a 2D induced density decays faster than the actual one, making the neighboring layers less effective at screening the electron-hole interaction. Second, the dipole response of the layers, which would increase the screening even more, is not included. In particular we mention that within the Q2D approach, removing the dipole contribution increases the converged value of the binding energy by $0.07$eV.
To conclude this paragraph, we have shown that even though the 2D model does capture the essential non-linear shape of $\epsilon(q_\parallel)$ in the small $q_\parallel$ regime, it underestimates the effect of environmental screening and consequently predicts too small changes in exciton binding energies due to substrate effects. 

\subsection{Transition from 2D to 3D Excitons in MoS$_2$}
\begin{figure}[t]
\centering  
\includegraphics[width=0.45\textwidth]{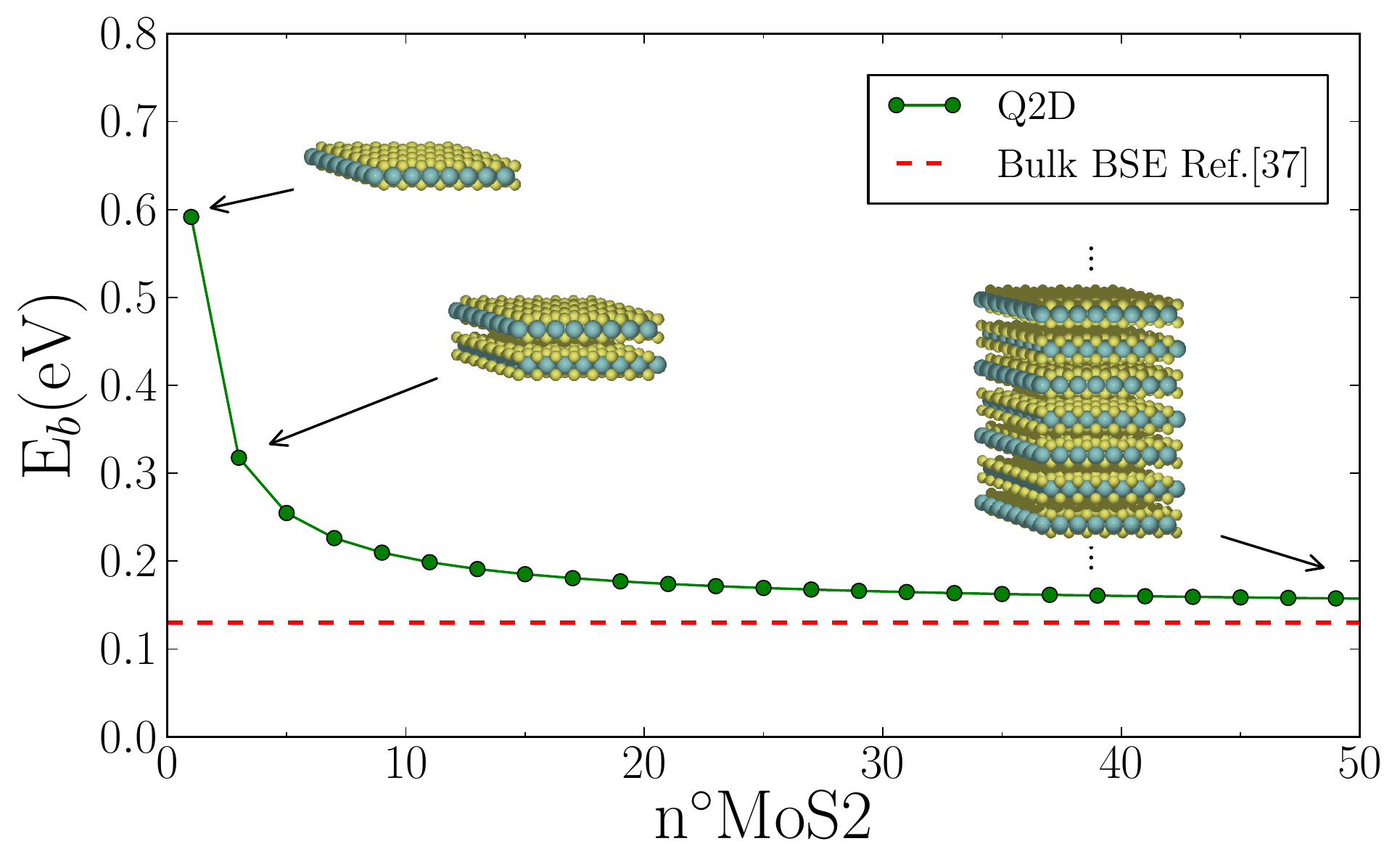}
\caption{Energy of the lowest bound exciton for MoS$_2$ incapsulated in MoS$_2$ layers as function of the total number of MoS$_2$ layers obtained from the Q2D (green) and 2D (blue) approaches. With the Q2D approach we can clearly see the transition from the mono-layer exciton to bulk one.}
\label{fig:VdW_Eb_MoS2}
\end{figure}

As a final example, we study the 2D to 3D transition of the exciton in MoS$_2$. In layered bulk materials, the Mott-Wannier equation can be written as follow:
\begin{equation}
\label{eq:3DMWHamiltonian}
\left[-\frac{\nabla_{\parallel}^2}{2\mu^{ex}_{\parallel}}-\frac{\nabla_{\perp}^2}{2\mu^{ex}_{\perp}}+W({\bf r})\right]F({\bf r})=E_bF({\bf r}),
\end{equation}
where typically the exciton mass in the out of plane direction is much higher than the in-plane directions ($\mu^{ex}_{\perp}\gg \mu^{ex}_{\parallel}$). Consequently, we can neglect the out-of-plane component of the kinetic energy and be left with the 2D Mott-Wannier model. Additionally, the in-plane effective mass does not vary considerably going from monolayer to bulk MoS$_2$ as shown in Ref. \onlinecite{Peelaers2012}. Therefore, the main difference between the physics of excitons in monolayer and layered bulk is contained in the screened potential rather than the geometric confinement. Based on this, it is tempting to model the bulk exciton as an electron-hole pair confined to a single layer but with an interaction screened by the bulk environment. To test this we consider a multi-layer MoS$_2$ structure and calculate the binding energy of an exciton localized in the central MoS$_2$ layer using the Q2D Mott-Wannier model with screened potential calculated from the QEH model. The results for the binding energy as a function of the number of MoS$_2$ layers are plotted in \cref{fig:VdW_Eb_MoS2}. 
As expected, the reduction of the exciton binding energy is larger when the monolayer is embedded in MoS$_2$ than in the case of hBN (\cref{fig:VdW_Eb} panel(b)). Amazingly, the binding energy converges towards a value of $0.16eV$ only $0.03eV$ higher than previously reported \emph{ab-initio} value for bull MoS$_2$\cite{Komsa2012}. This shows that the different nature of excitons in 2D and layered 3D materials is mainly caused by the screening while quantum confinement plays a minor role. 

\section{Conclusions}
In this work we have presented a systematic study of the screening properties of two-dimensional semiconductors and layered structures. Taking into account the finite extension of the 2D material in the out-of-plane direction, we have proposed a general quasi-2D picture to describe the screened electron-hole interaction in the context of excitons.  We have shown that, in the case of isolated layers, the excitons are typically large enough that the screening can be described by a linear dielectric function consistent with a strict 2D picture. On the other hand, for multi-layer structures where the screening properties are intermediate between the 2D and 3D regimes, it is essential to include the non-linear $q$-dependence of the dielectric function. If this is done and a quasi-2D description is employed, very satisfactory results are obtained for both monolayer and multi-layer structures using the same theoretical framework. In combination with a recently introduced scheme for computing dielectric functions of layered materials\cite{Andersen2015}, this makes it possible to model exciton physics in general van der Waals heterostructures at very low computational cost.  

\section{Acknowledgement}
The authors acknowledge support from the Danish Council for Independent Research's Sapere Aude Program through grant no. 11-1051390. The Center for Nanostructured Graphene (CNG) is sponsored by the Danish National Research Foundation, Project DNRF58.

\appendix
\section{Poisson's equation for lines of charge}
\label{appendix:Poisson}
Charges in 2D materials can be depicted as lines extending over the thickness of the layer. The potential generated by a line of charge can be calculated from the Poisson equation:
\begin{equation}
\label{eq:Peq}
\nabla^2 \varphi({\bf r})=-4\pi\rho({\bf r}).
\end{equation}
Because of the cylindrical symmetry of the line of charge, it is convenient to Fourier transform in the in-plane direction and rewrite  \cref{eq:Peq} as:
\begin{equation}
\label{eq:PeqFT}
\left[-|{\bf q}_\parallel|^2-\frac{\partial^2}{\partial z^2}\right]\varphi({\bf q}_\parallel,z)=-4\pi\rho({\bf q}_\parallel,z).
\end{equation}
For a line of charge, the density distribution can be separated as an in-plane delta function and an out-of-plane function $\rho(z)$ and therefore its in-plane Fourier transform would read $\rho({\bf q}_\parallel,z)=e^{-i{\bf q}_\parallel\cdot{\bf r}_\parallel}\rho(z)$. From the structure of \cref{eq:PeqFT} and the form of the Fourier transformed density, it is convenient to write the potential as $\varphi({\bf q}_\parallel,z)=\frac{e^{-i{\bf q}_\parallel\cdot{\bf r}_\parallel}}{|{\bf q}_\parallel|^2}\phi(z,{\bf q}_\parallel)$. Note that $\frac{e^{-i{\bf q}_\parallel\cdot{\bf r}_\parallel}}{|{\bf q}_\parallel|^2}$ is the Fourier transformed solution for the Poisson equation for a point charge in a 2D plane, therefore we can consider $\phi(z,{\bf q}_\parallel)$ as the out-of-plane component of the potential. Plugging $\varphi(z,{\bf q}_\parallel)$ and $\rho({\bf q}_\parallel,z)$ in \cref{eq:PeqFT}, we finally obtain the Poisson equation for the out-of-plane potential generated by a line of charge:
\begin{equation}
\label{eq:outofplanePot}
\frac{\partial^2}{\partial z^2}\phi(z,{\bf q}_\parallel)-|{\bf q}_\parallel|^2\phi(z,{\bf q}_\parallel)=-4\pi|{\bf q}_\parallel|^2\rho(z).
\end{equation}

\section{Unscreened Q2D Interaction}
\label{appendix:bareQ2D}
In this appendix we derive the expression for the Q2D unscreened charge-charge interaction in \cref{eq:bareVQ2D}. According to our Q2D picture and assuming a charge distribution $\rho_i({\bf r}_\parallel,z)=\frac{q_i\delta({\bf r}_\parallel-{\bf r}_{i,\parallel})}{d}\theta(\frac{d}{2}-|z-z_0|)$, the unscreened charge-charge interaction in reciprocal space can be written as:
\begin{equation}
\label{eq:Q2Dkspace}
\begin{split}
V_{Q2D}({\bf q}_\parallel) =& q_1q_2\int_V d{\bf r}\frac{\theta(\frac{d}{2}-|z-z_0|)e^{i{\bf q}_\parallel\cdot{\bf r}_{\parallel}}}{d} \times \\
&\int_Vd{\bf r}'\frac{1}{|{\bf r}-{\bf r}'|}\frac{\theta(\frac{d}{2}-|z'-z_0|)e^{-i{\bf q}_\parallel\cdot{\bf r}_{\parallel}'}}{d}.
\end{split}
\end{equation}
To proceed we notice that the integral in the second line can be interpreted as the potential generated by an in-plane Fourier transformed charge distribution $\rho({\bf q}_\parallel,z)=\frac{\theta(\frac{d}{2}-|z-z_0|)}{d}e^{-i{\bf q}_\parallel\cdot{\bf r}_{\parallel}}$ and its analytic form can be obtained solving  \cref{eq:PeqFT} as illustrated in \cref{appendix:Poisson}: 
\begin{equation}
\label{eq:potcharge}
\begin{split}
&\varphi({\bf q}_\parallel,z) = \frac{4\pi e^{-i{\bf q}_\parallel\cdot{\bf r}_{\parallel}}}{d|{\bf q}_\parallel |^2}\times\\
&\begin{cases}
1-e^{-|{\bf q}_\parallel |d/2}\cosh(|{\bf q}_\parallel | |z'-z_0|) & |z'-z_0|<\frac{d}{2} \\
e^{-|{\bf q}_\parallel | |z'-z_0|}\sinh(|{\bf q}_\parallel | d/2) & |z'-z_0|>\frac{d}{2}
\end{cases}.
\end{split}
\end{equation}
Plugging this result in \cref{eq:Q2Dkspace} and integrating in-plane and along z separately, we recover the expression \cref{eq:bareVQ2D}

\section{Screened Q2D Interaction}
\label{appendix:screenedQ2D}
In the following we show how to derive the expression for the Q2D screened electron-hole interaction reported in \cref{eq:WQ2D_q}.
For charge distributions of the kind $\rho_i({\bf r}_\parallel,z)=\frac{\delta({\bf r}_\parallel-{\bf r}_{i,\parallel})}{d}\theta(\frac{d}{2}-|z-z_0|)$, the screened interaction reads:
\begin{equation}
\label{eq:screenedQ2Dkspace}
\begin{split}
W_{Q2D}({\bf q}_\parallel) =& -\int_V d{\bf r}d{\bf r'}\frac{\theta(\frac{d}{2}-|z-z_0|)e^{i{\bf q}_\parallel\cdot{\bf r}_{\parallel}}}{d} \times \\
&\hspace{-1cm}\epsilon^{-1}({\bf r},{\bf r'})\int_Vd{\bf r}''\frac{1}{|{\bf r}'-{\bf r}''|}\frac{\theta(\frac{d}{2}-|z''-z_0|)e^{-i{\bf q}_\parallel\cdot{\bf r}_{\parallel}''}}{d}.
\end{split}
\end{equation}
As done in \cref{appendix:bareQ2D} we can interpret the integral in the second line as the potential in \cref{eq:potcharge}. In order to keep the calculation analytic we approximate $\varphi({\bf q}_\parallel,z)$ with its average inside the slab in the out-of-plane direction, in formula:
\begin{equation}
\label{eq:potchargeapprox}
\begin{split}
\varphi({\bf q}_\parallel,z) &\simeq \frac{1}{d}\int_{z_0-d/2}^{z_0+d/2} dz\varphi({\bf q}_\parallel,z)\\
&= -\frac{4\pi e^{-i{\bf q}_\parallel\cdot{\bf r}_{\parallel}}}{d^2|{\bf q}_\parallel |^2}\left(1-\frac{2}{|{\bf q}_\parallel|d}e^{-|{\bf q}_\parallel|d/2}\sinh({\bf q}_\parallel d/2)\right)\\
&= \frac{e^{-i{\bf q}_\parallel\cdot{\bf r}_{\parallel}}}{d}V_{Q2D}({\bf q}_\parallel)
\end{split}
\end{equation}
Inserting the last expression in \cref{eq:potchargeapprox} and integrating in-plane we get:
\begin{equation}
\begin{split}
W_{Q2D}({\bf q}_\parallel)&=V_{Q2D}({\bf q}_\parallel)\frac{1}{d}\int_{z_0-d/2}^{z_0+d/2}dz\int_{z_0-L/2}^{z_0+L/2}\epsilon_{00}^{-1}(z,z')\\
&=V_{Q2D}({\bf q}_\parallel)\epsilon_{Q2D}^{-1}({\bf q}_\parallel).
\end{split}
\end{equation}
\vspace{1cm}

\bibliographystyle{ieeetr}

\begin{thebibliography}{10}

\bibitem{Wang2012}
Q.~H. Wang, K.~Kalantar-Zadeh, A.~Kis, J.~N. Coleman, and M.~S. Strano,
  ``Electronics and optoelectronics of two-dimensional transition metal
  dichalcogenides,'' {\em Nature nanotechnology}, vol.~7, no.~11, pp.~699--712,
  2012.

\bibitem{Fai2010}
K.~F. Mak, C.~Lee, J.~Hone, J.~Shan, and T.~F. Heinz, ``Atomically thin
  MoS$_2$: A new direct-gap semiconductor,'' {\em Phys. Rev.
  Lett.}, vol.~105, p.~136805, Sep 2010.

\bibitem{Splendiani2010}
A.~Splendiani, L.~Sun, Y.~Zhang, T.~Li, J.~Kim, C.-Y. Chim, G.~Galli, and
  F.~Wang, ``Emerging photoluminescence in monolayer MoS$_2$,'' {\em Nano
  letters}, vol.~10, no.~4, pp.~1271--1275, 2010.

\bibitem{Ramasubramaniam2012}
A.~Ramasubramaniam, ``Large excitonic effects in monolayers of molybdenum and
  tungsten dichalcogenides,'' {\em Phys. Rev. B}, vol.~86, p.~115409, Sep 2012.

\bibitem{Qiu2013}
D.~Y. Qiu, F.~H. da~Jornada, and S.~G. Louie, ``Optical spectrum of
  MoS$_2$: Many-body effects and diversity of exciton states,''
  {\em Phys. Rev. Lett.}, vol.~111, p.~216805, Nov 2013.

\bibitem{Ugeda2014}
M.~M. Ugeda, A.~J. Bradley, S.-F. Shi, H.~Felipe, Y.~Zhang, D.~Y. Qiu, W.~Ruan,
  S.-K. Mo, Z.~Hussain, Z.-X. Shen, {\em et~al.}, ``Giant bandgap
  renormalization and excitonic effects in a monolayer transition metal
  dichalcogenide semiconductor,'' {\em Nature materials}, 2014.

\bibitem{Keliang2014}
K.~He, N.~Kumar, L.~Zhao, Z.~Wang, K.~F. Mak, H.~Zhao, and J.~Shan, ``Tightly
  bound excitons in monolayer WSe$_2$'' {\em Phys. Rev. Lett.},
  vol.~113, p.~026803, Jul 2014.

\bibitem{Jariwala2014}
D.~Jariwala, V.~K. Sangwan, L.~J. Lauhon, T.~J. Marks, and M.~C. Hersam,
  ``Emerging device applications for semiconducting two-dimensional transition
  metal dichalcogenides,'' {\em ACS nano}, vol.~8, no.~2, pp.~1102--1120, 2014.

\bibitem{Bernardi2013}
M.~Bernardi, M.~Palummo, and J.~C. Grossman, ``Extraordinary sunlight
  absorption and one nanometer thick photovoltaics using two-dimensional
  monolayer materials,'' {\em Nano letters}, vol.~13, no.~8, pp.~3664--3670,
  2013.

\bibitem{Lopez2013}
O.~Lopez-Sanchez, D.~Lembke, M.~Kayci, A.~Radenovic, and A.~Kis,
  ``Ultrasensitive photodetectors based on monolayer MoS$_2$,'' {\em Nature
  nanotechnology}, vol.~8, no.~7, pp.~497--501, 2013.

\bibitem{Ross2014}
J.~S. Ross, P.~Klement, A.~M. Jones, N.~J. Ghimire, J.~Yan, D.~Mandrus,
  T.~Taniguchi, K.~Watanabe, K.~Kitamura, W.~Yao, {\em et~al.}, ``Electrically
  tunable excitonic light-emitting diodes based on monolayer WSe$_2$ pn
  junctions,'' {\em Nature nanotechnology}, vol.~9, no.~4, pp.~268--272, 2014.

\bibitem{Pospischil2014}
A.~Pospischil, M.~M. Furchi, and T.~Mueller, ``Solar-energy conversion and
  light emission in an atomic monolayer pn diode,'' {\em Nature
  nanotechnology}, vol.~9, no.~4, pp.~257--261, 2014.

\bibitem{GrossoPastori2000}
G.~Grosso and G.~Parravicini, {\em Solid State Physics}.
\newblock Elsevier Science, 2000.

\bibitem{Cudazzo2010}
P.~Cudazzo, C.~Attaccalite, I.~V. Tokatly, and A.~Rubio, ``Strong
  charge-transfer excitonic effects and the bose-einstein exciton condensate in
  graphane,'' {\em Phys. Rev. Lett.}, vol.~104, p.~226804, Jun 2010.

\bibitem{Cudazzo2011}
P.~Cudazzo, I.~V. Tokatly, and A.~Rubio, ``Dielectric screening in
  two-dimensional insulators: Implications for excitonic and impurity states in
  graphane,'' {\em Phys. Rev. B}, vol.~84, p.~085406, Aug 2011.

\bibitem{Falco2013}
F.~H\"user, T.~Olsen, and K.~S. Thygesen, ``How dielectric screening in
  two-dimensional crystals affects the convergence of excited-state
  calculations: Monolayer MoS$_2$,'' {\em Phys. Rev. B}, vol.~88, p.~245309,
  Dec 2013.

\bibitem{Pulci2012}
O.~Pulci, P.~Gori, M.~Marsili, V.~Garbuio, R.~Del~Sole, and F.~Bechstedt,
  ``Strong excitons in novel two-dimensional crystals: Silicane and
  germanane,'' {\em EPL (Europhysics Letters)}, vol.~98, no.~3, p.~37004, 2012.

\bibitem{Hybertsen2013}
T.~C. Berkelbach, M.~S. Hybertsen, and D.~R. Reichman, ``Theory of neutral and
  charged excitons in monolayer transition metal dichalcogenides,'' {\em Phys.
  Rev. B}, vol.~88, p.~045318, Jul 2013.

\bibitem{Terrones2013}
H.~Terrones, F.~L{\'o}pez-Ur{\'\i}as, and M.~Terrones, ``Novel hetero-layered
  materials with tunable direct band gaps by sandwiching different metal
  disulfides and diselenides,'' {\em Scientific reports}, vol.~3, 2013.

\bibitem{Britnell2013}
L.~Britnell, R.~Ribeiro, A.~Eckmann, R.~Jalil, B.~Belle, A.~Mishchenko, Y.-J.
  Kim, R.~Gorbachev, T.~Georgiou, S.~Morozov, {\em et~al.}, ``Strong
  light-matter interactions in heterostructures of atomically thin films,''
  {\em Science}, vol.~340, no.~6138, pp.~1311--1314, 2013.

\bibitem{Geim2013}
A.~Geim and I.~Grigorieva, ``Van der waals heterostructures,'' {\em Nature},
  vol.~499, no.~7459, pp.~419--425, 2013.

\bibitem{Andersen2015}
K.~Andersen, S.~Latini, and K.~S. Thygesen, ``Dielectric genome of van der
  waals heterostructures,'' {\em Nano letters}, vol.~15, no.~7, pp.~4616--4621,
  2015.

\bibitem{Adler1962}
S.~L. Adler, ``Quantum theory of the dielectric constant in real solids,'' {\em
  Physical Review}, vol.~126, no.~2, p.~413, 1962.

\bibitem{Cheiwchanchamnangij2012}
T.~Cheiwchanchamnangij and W.~R.~L. Lambrecht, ``Quasiparticle band structure
  calculation of monolayer, bilayer, and bulk MoS$_2$,'' {\em Phys. Rev.
  B}, vol.~85, p.~205302, May 2012.

\bibitem{Molina2013}
A.~Molina-S\'anchez, D.~Sangalli, K.~Hummer, A.~Marini, and L.~Wirtz, ``Effect
  of spin-orbit interaction on the optical spectra of single-layer,
  double-layer, and bulk mos${}_{2}$,'' {\em Phys. Rev. B}, vol.~88, p.~045412,
  Jul 2013.

\bibitem{database2015}
``The dielectric building blocks and the qeh software can be downloaded
  from:.'' \url{https://cmr.fysik.dtu.dk/vdwh/vdwh.html}.
\newblock Accessed: 2015-08-18.

\bibitem{Filip2015}
F.~A. Rasmussen and K.~S. Thygesen, ``Computational 2D materials database:
  Electronic structure of transition metal dichalcogenides and oxides,'' {\em
  The Journal of Physical Chemistry C}, 2015.

\bibitem{Huser2013}
F.~H\"user, T.~Olsen, and K.~S. Thygesen, ``Quasiparticle gw calculations for
  solids, molecules, and two-dimensional materials,'' {\em Phys. Rev. B},
  vol.~87, p.~235132, Jun 2013.

\bibitem{Strinati1984}
G.~Strinati, ``Effects of dynamical screening on resonances at inner-shell
  thresholds in semiconductors,'' {\em Phys. Rev. B}, vol.~29, pp.~5718--5726,
  May 1984.

\bibitem{Onida2002}
G.~Onida, L.~Reining, and A.~Rubio, ``Electronic excitations:
  density-functional versus many-body green's-function approaches,'' {\em Rev.
  Mod. Phys.}, vol.~74, pp.~601--659, Jun 2002.

\bibitem{Rozzi2006}
C.~A. Rozzi, D.~Varsano, A.~Marini, E.~K.~U. Gross, and A.~Rubio, ``Exact
  coulomb cutoff technique for supercell calculations,'' {\em Phys. Rev. B},
  vol.~73, p.~205119, May 2006.

\bibitem{Enkovaara2010}
J.~Enkovaara, C.~Rostgaard, J.~J. Mortensen, J.~Chen, M.~Du{\l}ak, L.~Ferrighi,
  J.~Gavnholt, C.~Glinsvad, V.~Haikola, H.~Hansen, {\em et~al.}, ``Electronic
  structure calculations with gpaw: a real-space implementation of the
  projector augmented-wave method,'' {\em Journal of Physics: Condensed
  Matter}, vol.~22, no.~25, p.~253202, 2010.

\bibitem{GPAW_web}
``The {\bf gpaw} code is available as part of the campos software:.''
  \url{https://wiki.fysik.dtu.dk/gpaw/}.
\newblock Accessed: 2010-09-30.

\bibitem{Yan2012}
J.~Yan, K.~W. Jacobsen, and K.~S. Thygesen, ``Optical properties of bulk
  semiconductors and graphene/boron nitride: The bethe-salpeter equation with
  derivative discontinuity-corrected density functional energies,'' {\em Phys.
  Rev. B}, vol.~86, p.~045208, Jul 2012.

\bibitem{Chernikov2014}
A.~Chernikov, T.~C. Berkelbach, H.~M. Hill, A.~Rigosi, Y.~Li, O.~B. Aslan,
  D.~R. Reichman, M.~S. Hybertsen, and T.~F. Heinz, ``Exciton binding energy
  and nonhydrogenic rydberg series in monolayer WS$_2$,'' {\em
  Phys. Rev. Lett.}, vol.~113, p.~076802, Aug 2014.

\bibitem{Peelaers2012}
H.~Peelaers and C.~G. Van~de Walle, ``Effects of strain on band structure and
  effective masses in MoS$_2$,'' {\em Phys. Rev. B}, vol.~86, p.~241401,
  Dec 2012.

\bibitem{Komsa2012}
H.-P. Komsa and A.~V. Krasheninnikov, ``Effects of confinement and environment
  on the electronic structure and exciton binding energy of MoS$_2$ from
  first principles,'' {\em Phys. Rev. B}, vol.~86, p.~241201, Dec 2012.

\end{thebibliography}

\end{document}